\documentclass[a4paper,11pt] {ivoa} 
\usepackage{geometry}                % See geometry.pdf to learn the layout
\usepackage{graphicx}
\usepackage{amssymb}
\usepackage{epstopdf}
\usepackage{amsmath}
\usepackage{hyperref}
\usepackage{color}
\usepackage{xcolor}
\usepackage{listings}

\definecolor{gray}{rgb}{0.4,0.4,0.4}
\definecolor{darkblue}{rgb}{0.0,0.0,0.6}
\definecolor{cyan}{rgb}{0.0,0.6,0.6}

\lstdefinestyle{listXML}{language=XML, extendedchars=true,  belowcaptionskip=5pt, xleftmargin=1.8em, xrightmargin=0.5em, numbers=left, numberstyle=\small\ttfamily\bf, 
frame=single, breaklines=true, breakatwhitespace=true, breakindent=0pt, emph={}, emphstyle=\color{red}, basicstyle=\small\ttfamily,
columns=fullflexible, showstringspaces=false, commentstyle=\color{gray}\upshape,
morestring=[b]",
morecomment=[s]{<?}{?>},
morecomment=[s][\color{orange}]{<!--}{-->},
keywordstyle=\color{cyan},
stringstyle=\color{black},
tagstyle=\color{darkblue},
morekeywords={xmlns,version,type}
}

\title{PDL: The Parameter Description Language}
\editor{Carlo Maria Zw\"olf }
\author{Carlo Maria Zw\"olf \\ Paul Harrison \\ Juli\'an Garrido \\ Jose Enrique Ruiz \\ Franck Le Petit}
\newcommand{\pdlversion}{1.0}
\newcommand{\pdldate}{May 23, 2014}

\date{\pdldate}

\ivoatype{IVOA Recommendation}
\ivoagroup{Grid and Web Services}
\version{\pdlversion}
\def\SVN$#1: #2 ${\expandafter\def\csname SVN#1\endcsname{#2}}
\SVN$Rev: 201 $

\urlthisversion{\pdlversion:  \pdldate\ }
\urllastversion{N/A}
\previousversion{PDL: The Parameter Description Language, IVOA Proposed Recommendation, Version 1.0
}

\begin{document}
\maketitle
\section*{Abstract}
This document discusses the definition of the  {\it Parameter Description Language} (PDL). In this language parameters are described in a rigorous data model. With no loss of generality, we will represent this data model using XML.\\
It intends to be a expressive language for self-descriptive web services exposing the semantic nature of input and output parameters, as well as all necessary complex constraints. PDL is a step forward towards true web services interoperability. 
\section{Status of this document}
This document has been produced by the Grid and Web Service Working Group. It follows the previous working draft.
 
\section*{Acknowledgements}
We wish to thank the members of the IVOA Grid and Web Services working group for the discussions around PDL it has hosted during the Interop meetings (starting from Naples, May 2011). A special thanks are due to Andr\'e Schaaff for his advice, useful discussions and feedback on every version of this work.

\clearpage

\tableofcontents

\newpage

\section{Preface}
Before going into technical details and explanations about PDL, we would like to suggest what are the categories of users potentially interested in this description language and underline in what field PDL has a strategic impact.\\

 PDL is particularly addressed to scientists or engineers
\begin{itemize}
\item wishing to expose their research codes (without limits or compromise on the complexity of the exposed code) online as public services,
\item wishing to interconnect their codes into workflows.
\end{itemize}
We will continue this preface by a quick `theoretical' overview of PDL. For a practice-oriented introduction, reader should refer to the annex paragraph \ref{divePDL}.\\

 {\bf The online code aspect -} Usually, people who are about to publish their codes as online services are wondering if user will be able to correctly use the new services: the code may have many parameters (usually with obscure names,  due to historical reasons). These parameters can be intercorrelated and/or have forbidden ranges, according to the configuration or validity domain of the implemented model. In other words the use of the code may be reserved to experts.\\
How can the service provider ensure that the users will be aware of all the subtleties (units, physical meaning, value ranges of parameters)? The natural answer is to provide a good code documentation to the community. However, our experience as service providers showed that rarely users read carefully the documentation. And even if they read it entirely, they are not safe from making gross and/or distraction errors. 
Two commons consequences of this situation are the abandonment of the service by users, after few inconclusive tests and
the abandonment of the service by the provider him(her)self, buried by e-mails from users containing question whose answer are ... in the provided documentation.\\
PDL is a powerful tool for improving both the user and the provider experiences: it may be seen as a way for hardcoding all the subtleties and constraints of the code into the software components used by the provider for exposing the code and by users for interacting with it.\\
The fine expertise on the code to expose is fixed into the PDL description. The PDL software framework is able to automatically generate client interfaces (for assisting the user in interacting with the service) and checking algorithms (for verifying that the data submitted by users are compliant with the description). Moreover the softwares composing the PDL framework are generic elements which are automatically configured by the description into ad-hoc components (cf. paragraph \ref{SoftwareImplementation} for further details and explanation about these concepts). The work for people wishing to expose code is indeed essentially restricted to redaction of the description. For these reasons PDL is particularly indicated for people wishing to expose their code but  don't have much time or the technical skills for building web services.\\

 {\bf The workflow aspect -} Scientists or engineers wishing to integrate their code into workflow engines have to write ad hoc artifacts: (in the case of the {\it Taverna} engine \cite{Taverna2}) these could be {\it Java Beans}, {\it Shell} and {\it Python} artefacts. Normally one has to write a specific artefact for every code to integrate. This could be extremely time consuming and, from the technical point of view, this is not an `out of the box' procedure for people starting using workflows engine.\\
PDL is indicated to facilitate the integration of code into workflow engines: the integration phase is reduced to the redaction of the description.\\ 
Moreover PDL introduce a new feature into the workflow domain: since every description embeds fine grained details and metadata on parameters (also with their related constraints), the physical sense (meaning and integrity) of a workflow could be automatically verified.\\

The PDL application domain is not limited only to online code or workflows. We are now going to detail all the technical aspects of this description grammar.

\section{Introduction}
In the context of the {\it International Virtual Observatory Alliance} researchers would like to
provide astronomical services to the community. \\
These services could be 
\begin{itemize}
\item   access to an existing catalogue of images and/or data,
\item  access to smaller sub-products images, spectra and/or data generated on the fly,
\item  the entry point to a database listing the results of complex and compute-intensive numerical simulations,
\item a computation code exposed online, etc... 
\end{itemize}
In the following we will ignore any specific feature and will use the term {\it generic service} to
refer to any kind of process that receives input parameters and produces output ones.\\

Interoperability with other services and the immediacy of use are two key factors in the success of a service:
in general, a service will not be used by the community if users do not know how to call it, the inputs it needs, or what it does and how. However, other issues may have influence in the user decision e.g. Who has implemented it? Who is the service provider? Does it implement a well known technique? Is there a paper to support the research behind the service? Can it be used as a standalone application and can it be used together with other services? 
A new service will be more useful for some users if it can be released easily as an interactive and standalone application whereas for other users the interoperability with other services and applications is more important. This standard is focused on the needs of the second group, as the ease of the distribution of web services is the primary concern of service providers.
Indeed, service description and interoperability are two key points for building efficient and useful ecosystem of services. 

PDL aims to provide a solution to the problems of description and interoperability of services. 
With PDL, service providers will be able to share with users (either as humans or as computer systems) 
the knowledge of what the service does (and how).
Moreover this new service will be immediately interactive and well integrated with other services.\\

{\bf Service description} and {\bf Interoperability} are indeed two key points for building
efficient and useful services.

\subsection{The service description: existing solutions and specific needs}
For a client starting to interact with an unknown service, its description is fundamental: in a
sense it is this description that puts the service from the {\it unknown} to the {\it known}
state.\\
Since the client could be a computer system, a generic description should be machine-readable.\\

There are several pre-existing service description languages.
 The most well known for their high expression level and their
wide use are the \emph{W3C} \emph{WSDL} and
\textit{WADL} \cite{WSDL}, \cite{WADL}.\\
Since both {\it WSDL} and {\it WADL} support {\it XML-schema}, one could include in these descriptions complex and highly specialized 
XML objects for expressing conditions and/or restrictions. However, the process for building these ad-hoc XML extension types is not standard\footnote{For example, for expressing that a given parameter must be greater and smaller than arbitrary values, we could define a {\it bounded} type containing an {\it inf} field and a {\it sup} field. If another user defines a similar object calling these two fields {\it inf-bound} and {\it sub-bound}, the instances of these two types could not interoperate straightforwardly. The theory of types is not sufficient to ensure the interoperability of the object defined.}: 
a service provider could only describe, using the native standard feature of WADL or WSDL, primitive-typed parameters. It thus serves a roughly similar purpose as a method-signature in a programming language, with no possibility for defining 
restrictions, semantics and criteria to satisfy. PDL proposes a way for expressing these features in a unified way.\\

In the case of {\it generic services} for science, the description needs are very specific: since we
have to deal with complex formalisms and models, one should be able to describe for each parameter; its
physical meaning, its unit and precision and a range (or set) of admissible values (according to
a model).\\ 
In many cases, especially for theoretical simulations, parameters could be linked by
complex conditions or have to satisfy, under given conditions, a set of constraints (that could
involve mathematical properties and formulas).

Two examples of this high level description we would be able to provide are the following:

\begin{equation}\label{PDLExemplum01}
\mbox{Service1 }\left\{
\begin{array}{l}
\ \mbox{Input } \left\{
\begin{array}{c}
 \mbox{$\vec p_1$ is a $m/s$ vector speed and $\| \vec p_1\|<c$} \\
 \mbox{ $p_2$ is time (in second) and $p_2 \geq 0$ }\\
 \mbox{$p_3$ is a $kg$ mass and $p_3 > 0$}\\
\end{array}
\right. \\
\\
\ \mbox{\hspace{1cm} Output } \left\{
\begin{array}{l}
 \mbox{ $p_4$ is a Joule Kinetic Energy and $p_4 \geq 0$} \\
 \mbox{ $p_5$ is a distance (in meter) }\\
 \end{array}
\right.\\
\end{array}
\right.
\end{equation}

\begin{equation}\label{PDLExemplum02}
\mbox{Service2 } \left\{
\begin{array}{l}
\ \mbox{Input } \left\{
\begin{array}{l}
\ \mbox{ $\mathbb R \ni p_1 >0$; $p_2 \in \mathbb N$; $p_3 \in \mathbb R$} \\
\  \bullet \mbox{ if $p_1 \in ]0,\pi/2]$ then $p_2 \in \{2;4;6\}$,}\\
\ \mbox{$p_3 \in [-1,+1]$ and $\displaystyle \left( \left|  \sin(p_1)^{p_2} -p_3 \right| \right)^{1/2}<3/2$ } \\
\ \bullet \mbox{ if $p_1 \in ]\pi/2,\pi]$ then $0<p_2 < 10$,}\\
\ \mbox{$p_3>\log(p_2)$ and $(p_1 \cdot p_2)$ must belong to $\mathbb N$} \\
\end{array}
\right. \\
\\
\ \mbox{\hspace{1cm} Output } \left\{
\begin{array}{l}
 \mbox{$\vec p_4, \, \vec p_5 \in \mathbb R^3$ } \\
 \ \mbox{Always $\displaystyle \frac{\| \vec p_5\|}{\|\vec p_4 \|} \leq 0.01 $} \\
 \end{array}
\right.\\
\end{array}
\right.
\end{equation}
To our knowledge, no existing description language meets these exacting requirements of scientific
services. This leads us naturally to work on a new solution and consider developing a new
description language.\\

 {\bf Remark: } The PDL descriptions for the two examples above are provided respectively in paragraphs \ref{Exemplum1XML}
and \ref{Exemplum2XML}.

\subsection{Interoperability issues}\label{ParInteropIssues}
Nowadays, with the massive spread and popularity of {\it cloud} services, interoperability has become
an important element for the success and usability of services. This remains true in the context of
astronomy.
For the astronomical community, the ability of systems to work together without restrictions (and
without further {\it ad hoc} implementations) is of high value: this is the ultimate goal
that guides the {\it IVOA}.\\

Computer scientists have developed different tools for setting up service interoperability and
orchestration. The most well known are
\begin{itemize}
\item {\it BAbel} \cite{Babel1}, \cite{Babel2}, \cite{Babel3} (\href{https://computation.llnl.gov/casc/components/}{https://computation.llnl.gov/casc/components/}),
\item {\it Taverna} \cite{Taverna1}, \cite{Taverna2} (\href{http://www.taverna.org.uk}{http://www.taverna.org.uk}),
\item {\it OSGI} and {\it D-OSGI } \cite{Osgi1} (\href{http://www.osgi.org/}{http://www.osgi.org/}),
\item {\it OPalm} \cite{Opalm1}, \cite{Opalm2}, \cite{Opalm3} (\href{http://www.cerfacs.fr/globc/PALM_WEB/}{http://www.cerfacs.fr/globc/PALM\_WEB/}),
\item {\it GumTree} \cite{GumTree1}, \cite{GumTree2} (\href{http://docs.codehaus.org/display/GUMTREE/}{http://docs.codehaus.org/display/GUMTREE/}).
\end{itemize}
In general, with those tools one could coordinate only the services written with given languages.
Moreover the interoperability is achieved only in a basic `computer' way: if the input of the $B$
service is a double and the output of the $A$ service is a double too, thus the two services could
interact.\\

Our needs are more complex than this: let us consider a service $B'$ whose inputs are a density and
a temperature and a service $A'$ whose outputs are density and temperature too. \\
The interoperability is not so straightforward: the interaction of the two services has a sense only
if the two densities (likewise the two temperatures)
\begin{itemize}
\item have the same `computer' type (ex. double),
\item are expressed in the same system of units,
\item correspond to the same physical concepts (for example, in the service $A'$ density could be
an electronic density whereas in the service $B'$ the density could be a mass density)
\end{itemize}
But things could be more complicated, even if all the previous items are satisfied: the model behind
the service $B'$ could implement an Equation of State which is valid only if the product
(density$\times$temperature) is smaller than a given value.
Thus the interoperability with $A'$ could be achieved only if the outputs of this last satisfy the
condition on product.\\

Again, as in case of descriptions no existing solutions could meet our needs and we are oriented
towards building our own solution.\\

 {\bf Remark}: We will present further considerations on the workflows aspects in paragraph \ref{PDLWF}, once we have exposed some basic concepts about PDL in the following paragraph.  

\subsection{Astronomical and Astrophysical use-cases needing PDL's fine description capabilities}
PDL was originally designed for meeting requirements coming from the community members wishing to expose their code online as public services. One of the difficulty they often mentioned is that online codes are often complex to use and users may do mistake with online simulations. For example, they could use them outside of their validity domain. The description of parameters with PDL allows to constrain those ones in validity domains, and so PDL answers this fear of the theorist community.\\

In order to build a grammar like PDL, we could go two ways: in the first we would have built a monolithic solution for meeting the vast majority of astronomical and astrophysical needs. In the other we would have to provide community with a flexible enough tool  (modular and extensible) to fit the majority of use-cases: if the parameters (of a given service) are decomposed with the finest granularity, PDL is a good tool for performing {\it a priori verification}, notifying errors to user before submitting jobs to a server system.
This has, for example, an immediate consequence on how we deal, in PDL, with sky coordinates: we don't have particular fields/entries for ascensions and declinations. For us this parameters could be stored in {\it double} parameters. The associated unit will precise if the angle will be considered in degrees or radians  and the associated {\it SKOS} concepts \cite{Skos1}, \cite{Skos2}
(\href{http://www.w3.org/TR/skos-reference/}{http://www.w3.org/TR/skos-reference/}) will provide further information. If a service provider has to define particular conditions on the angular distance between two coordinates $(asc_1, \, dec_1)$ and $(asc_2, \, dec_2)$ (e.g. $|asc_1- asc_2|+|dec_1-dec_2|<\epsilon$) he/she may use the expression capabilities of PDL (cf. paragraph \ref{par02})\\

 During the PDL development, close cooperation naturally born with the Workflow community. PDL indeed allow the real {\it scientific} interoperability (not only based on computer types) required by the Astronomy and Astrophysics workflow community.\\

 The following sections of this document could seems complex at first reading. This is because we present all the features and the descriptive richness  of PDL. Nevertheless this does  not mean that all PDL descriptions are necessarily complex. They could be complex in case of services with many parameters linked by many constraints. But PDL description could be very simple in case of simple services. 
For example the PDL description associated with a common cone search service is very simple. It could be consulted at the following URL:\\
 \href{
http://www.myexperiment.org/files/999/versions/4/download/AMIGA-PDL-Description.xml
}{http://www.myexperiment.org/files/999/versions/4/\\download/AMIGA-PDL-Description.xml}.

\subsection{A new Parameter Description Language: a unique solution to description and interoperability needs}\label{ANewPDL}
To overcome the lack of a solution to our description and interoperability
needs, it is proposed to introduce a new language.
Our aim is to finely describe the set of parameters (inputs and outputs of a given generic services)
in a way that
\begin{itemize}
\item could be {\it interpreted} by human beings (we could say {\it understood} for the simpler description cases),
\item could be parsed and handled by a computer,
\item complex relations and constraints involving parameters could be formulated unambiguously.
Indeed we would like to express
\begin{itemize}
\item mathematical laws/formulas,
\item conditional relationships (provided they have a logical sense)
\end{itemize}
involving parameters.
\end{itemize}
The new language is based on a generic data model (DM). Each object of the DM corresponds to a
syntactic element. Sentences are made by building object-structures.
Each sentence can be interpreted by a computer by parsing the object structure.\\

With PDL one could build a mathematical expression (respectively conditional sentences) assembling the base-element described in section \ref{par02} (resp. section \ref{complexRelations}). 

If a particular expression (or condition) could not be expressed using the existing features,
this modular grammar could be extended by introducing an ad hoc syntactic element into the object DM. \\

For describing the physical scientific concept or model behind a given parameter, the idea is to use
{\it SKOS} concepts and, if more complexity is required by the use case, a richer ontology \cite{Ontology}.\\

Since the inputs and outputs  of every service (including their constraints and complex conditions)
could be described with this fine grained granularity, interoperability becomes possible in the {\it
smart} and {\it intelligent} sense we really need: services should be able to work out if they can
reasonably use their output as input for another one, by simply looking at its description.\\

With no loss of generality and to ensure that the model could work with the largest possible number
of programming languages, we decided to fix it under the form of an XML schema (cf paragraph \ref{pdlSchema}).This choice is also
convenient because there are many libraries and tools for handling and parsing XML documents.\\

{\bf Remark:} We recall that PDL is a syntactic framework for describing parameters (with related
constraints) of generic services. Since a PDL description is rigorous and unambiguous, 
it is possible to verify if the instance of a given parameter (i.e. the value of the parameter
that a user sends to the service) is consistent with the description.\\
In what follows in this document, we will often use the terms {\it evaluate} and {\it interpret}
with reference to an expression and/or condition composed with PDL. By this we mean that one must
replace the referenced parameters (in the PDL expressions/conditions) by the set of values provided to the
service by the user. The replacement mechanisms will be explained in detail, case by case.

\subsubsection{PDL in the IVOA architecture}\label{IVOAarch}
\begin{figure}[htbp]
\begin{center}
\includegraphics[width=1.0\textwidth]{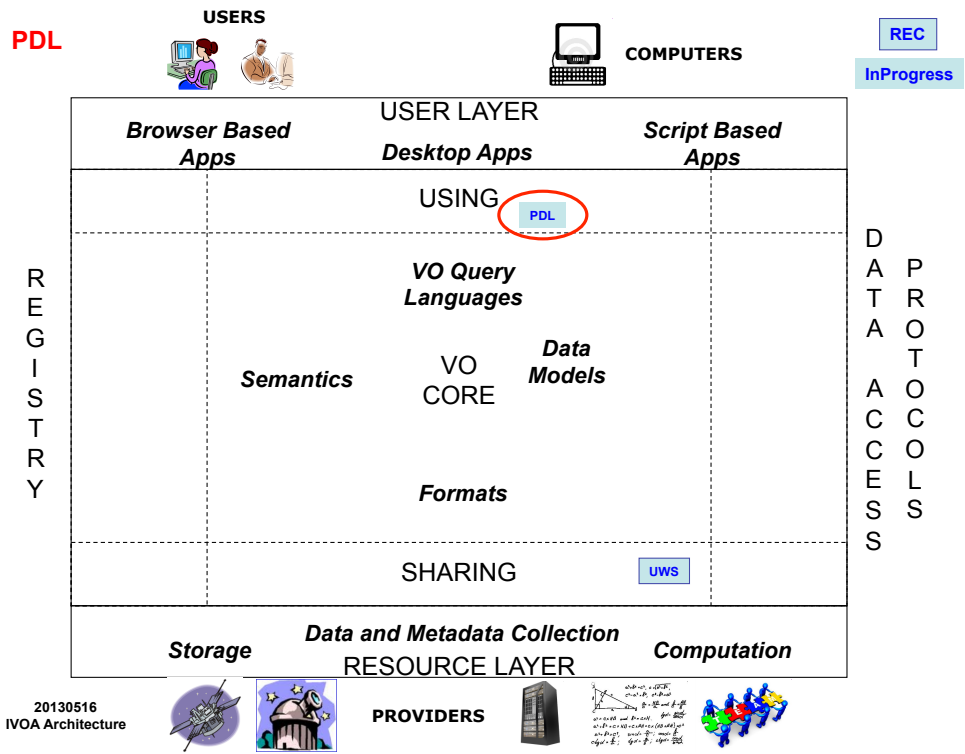} 
\caption{{\it The IVOA Architecture, with PDL highlighted. As pointed out in paragraph \ref{IVOAarch}, the domains and scopes of PDL and UWS are well separated: one can be used without the other without infringing any rules of those  standards. Of course they could work in synergy. In this case PDL could be seen a supplementary layer (explaining the physical/computational meaning of every parameter), whereas UWS has only a description of the values of parameters.}}
\label{Pic-arch}
\end{center}
\end{figure}
Within the IVOA Architecture of figure \ref{Pic-arch}, PDL is a VO standard for richly describing parameters with a fine grained granularity, allowing to introduce constraints and mathematical formulae.\\
If PDL describes the nature, the hierarchy of parameters and their constraints, {\bf it  does not describe} how this parameters are transmitted to a service, nor how these parameters will be processed by the described service. For example, PDL does not prescribe whether to transfer parameters through a SOAP envelope or through a REST post, nor what will be the  phases that the submitted job will pass through. In the context of the IVOA, this means that the separation between PDL and UWS \cite{UWS} is clear and one can be used without the other without infringing any rules of those  standards.\\
Indeed, PDL could be seen a supplementary layer, for explaining in a unified way the physical/computational meaning of every parameter, whereas UWS has only a description of the values of parameters. \\
PDL could be plugged as an additional layer to every existing IVOA service and is suitable for solving issues not covered by other IVOA standards and is particularly indicated for workflows.

\subsubsection{Some consideration on PDL and Workflows}\label{PDLWF}
The orchestration of services defines a Scientific Workflow, and services interoperability is key in the process of designing and building workflows. An important consideration in this process of orchestration is the control of parameters constraints at the moment of the workflow execution. Even if interoperability is assured at the phase of workflow design, a control at the execution phase has to be implemented by workflow engines as service clients. 
As we suggested in the remark of the previous paragraph, testing for valid parameters provided to a service could be automatically generated starting from the PDL description. 
%(cf for example the implementing note document REFERENCE to add after having it published on the IVOA site). 
This automation facility could be used to perform the verification on both client side and on server side:
\begin{itemize}
\item verifications made on client-side will avoid sending the wrong set of parameters to a server, reducing the load on the latter,
\item verifications made on server-side will avoid running jobs with wrong set of parameters. Indeed a server does not know if the job is sent by a client implementing the verifications or not. Therefore it must behave as if the data had never been checked. 
\end{itemize}
Verification of non-standard errors (e.g. network issues) are out of the scope of PDL.\\

\begin{figure}[htbp]
\begin{center}
\includegraphics[width=0.4\textwidth]{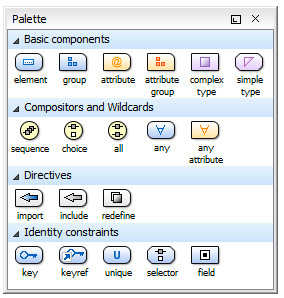} 
\caption{Graphical convention adopted for building graphical representations, starting from the XML schema.}
\label{figSchema}
\end{center}
\end{figure}

\subsubsection{On the graphical representations adopted into this document}
As recalled at the end of the paragraph \ref{ANewPDL}, we decided to fix the PDL grammar into an XML schema. The graphical diagrams proposed into this document are a simple rendering of every XML element contained into the schema, obtained following the graphical convention of the figure \ref{figSchema}.\\
Indeed, a list with the defined schema components (elements, attributes, simple and complex types, groups and attribute groups) is presented into the graphical representation: every complex element described is linked with segments to the contained sub-elements. A bold segment indicates that the sub-element is required and a thin segment indicates that the sub-element is optional. Moreover, the cardinality of the contained sub-elements could be expressed on the segments.

\section{The Service Class}\label{par-service}

\begin{figure}[htbp]
\begin{center}
\includegraphics[width=0.9\textwidth]{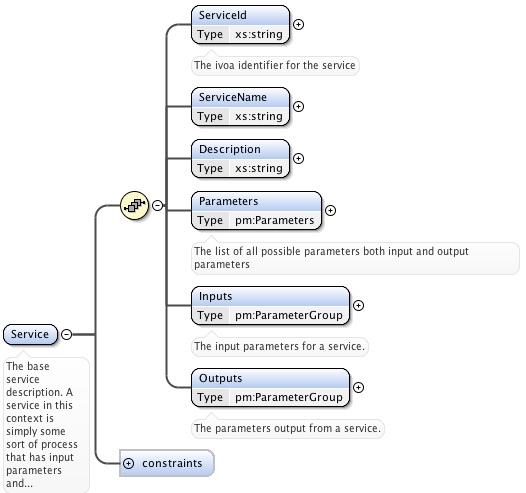} 
\caption{Graphical representation of the Service Class}
\label{Pic-Service}
\end{center}
\end{figure}

The root element of the PDL description of a generic service is the object {\it Service} (see figure
\ref{Pic-Service}). This {\bf must contain}
\begin{itemize}
\item A single {\it ServiceName}. This field is a String containing the name of the service.
\item A  {\it ServiceId}. This field is a String containing the IVOA id of the service.
It is introduced for a future  integration of PDL into the registries: each service in
the registry will be marked with its own unique id.
\item A  {\it Description}. This field is a String and contains a human readable description
of the service. This description is not intended to be understood/parsed by a machine.
\item A {\it Parameters} field which is a list of {\it SingleParameter} object types (cf.
paragraph \ref{par01}). This list contains the definition of all parameters (both inputs and
outputs) of the service. The two following fields specify if a given parameter is a input or an
output one.
\item An {\it Inputs} field of type {\it ParameterGroup} (cf. paragraph \ref{par-group}). This
object contains the detailed description (with constraints and conditions) of all the input
parameters.
\item An {\it Outputs} field of type {\it ParameterGroup}. This object contains the detailed
description (with constraints and conditions) of all the output parameters.
\end{itemize}

\section{The SingleParameter Class}\label{par01}

\begin{figure}[htbp]
\begin{center}
\includegraphics[width=0.65\textwidth]{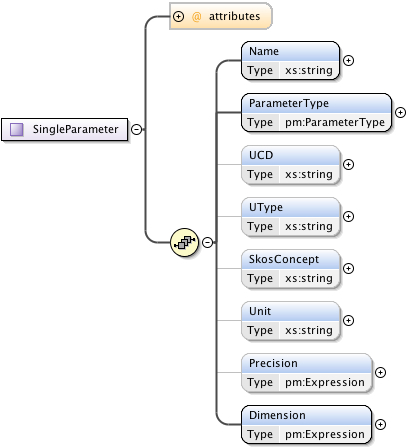} 
\caption{Graphical representation of the Parameter Class}
\label{Pic-Parameter}
\end{center}
\end{figure}

The {\it SingleParameter} Class (see figure \ref{Pic-Parameter}) is the core element for describing
jobs.
Every object of this type must be characterized by:
\begin{itemize}
\item A name, which is the Id of the parameter. In a given PDL description instance, two parameters cannot 
have the same name;
\item A single parameter type, which explains the nature of the current parameter. The allowed
types are : boolean, string, integer, real, date;
\item A dimension. A $1$-dimension corresponds to a scalar parameter whereas a dimension
equal to N corresponds to a N-size vector. The dimension is expressed using an {\it expression} (cf.
paragraph \ref{par02}). The result of the expression that appears in this {\it
SingleParameter}-field object {\bf must be integer}.\footnote{This is obvious, since this value
corresponds to a vector size.}
\end{itemize}

 {\bf Remark on the vector aspect:} It could seem unnecessarily complex to have the parameter dimension into an {\it Expression}. This feature has been introduced for meeting some particular needs: consider for example a service computing polynomial interpolations. Let the first parameter $d$ (an integer) be the degree of the interpolation and the second parameter  $\vec p$ be the vector containing the set of points to interpolate. For basic mathematical reasons, these two parameters are linked by the condition $(\mbox{size}(\vec v)-1) = d$. By defining dimensions as {\it Expressions} we can easily include this kind of constraints into PDL descriptions.\\
Vectors in PDL are intended as one dimensional arrays of values. Further significations should be documented using the UCD, Utype or the Skos concept fields. Moreover, if one wish to define an {\it Expression} using the scalar product of two vectors (cf. paragraph \ref{par02}) he/she has to pay attention that the involved vectors are expressed in the same orthonormal basis.\\

The attribute {\it dependency} can take one of the two values {\bf required} or {\bf
optional}. If required the parameter {\bf must be} provided to the service. If optional, the service
could work even without the current parameter and the values will be considered for processing only
if provided.\\

%proffreadend
Optional fields for the {\it SingleParameter} Class are:
\begin{itemize}
\item a UCD : which is a text reference to an existing UCD for characterizing the parameter \cite{UCD};
\item a Utype  : which is a reference to an existing Utype for characterizing the parameter \cite{Utype}
(\textcolor{red}{the reference is typically a text string});
\item a Skos Concept  (\textcolor{red}{the reference is typically a text string} containing the valid URL of a Skos concept).
\item a Unit (\textcolor{red}{which is a string reference to a valid VOUnits element}).
\item a precision. This field must be specified only for parameter types where the concept of
precision makes sense. It has indeed no meaning for integer, rational or string. It is valid, for instance, for a real type. To understand the meaning of this field, let the function
$f$ be a model of a given service. If $i$ denotes the input parameter, $f(i)$ denotes the output. The
precision $\delta$ is the smaller value such that $f(i+\delta) \neq f(i)$.\\ The precision is
expressed using an {\it expression} (cf. paragraph \ref{par02}). The result of the expression that
appears in this {\it precision}-field  {\bf must be} of the same type as (or could be naturally
cast to) the type appearing in the field {\it parameter type}.
%\item an optional restriction on the object which takes the form of an {\it
%AlwaysCondtionalStatement} (cf. paragraph \ref{par-AlwaysConditionalStatement}). This restriction is
%a way of placing a global restriction on a parameter that takes precedence over any constraint that
%might occur on a parameter in a {\it ParameterGroup}.
\end{itemize}

{\bf NB:} The name of every {\it SingleParameter} is unique. 

\section{The ParameterRef Class}\label{par-parRef}

\begin{figure}[htbp]
\begin{center}
\includegraphics[width=0.65\textwidth]{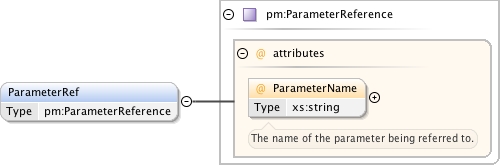} 
\caption{Graphical representation of the Parameter Reference Class}
\label{Pic-ParameterRef}
\end{center}
\end{figure}

This Class, as it name suggests, is used to reference an existing parameter defined in the {\it
Service} context (cf. paragraph
\ref{par-service}). It  contains only an attribute {\it ParameterName} of type String which
must corresponds to the {\it Name} field of an existing {\it SingleParameter} (cf. paragraph \ref {par01}).

\section{The ParameterType Class}\label{par-ParameterType}
This Class is used to explain the type of a parameter (cf. paragraph \ref{par01}) or an expression
(cf. paragraph \ref{par02_03}). The allowed types are : 
\begin{itemize}
\item Boolean. The allowed values for parameters of this type are {\it true / false}, non case sensitive.
\item String. Any String (UTF8 encoding recommend).
\item Integer. Numbers (positive and negatives) composed of [0-9] characters. 
\item Real. Two formats are allowed for parameters of this type: 
\begin{itemize}
\item numbers (positives and negatives) composed of [0-9] characters, with dot as decimal separator, 
\item scientific notation: number composed of [0-9] characters, with dot as decimal separator, followed by the {\it E} character (non case sensitive), followed by an integer. 
\end{itemize}
\item Date. Parameters of this type are dates in ISO8601 format.
\end{itemize}
{\bf Remark} There is a lot of complexity in expressing Date/Time values in astronomy. A first solution for covering all the  cases would be to include all the possibility into the data model. However, this hardcoded approach does not fit with the PDL modular approach. For example, if a service provider wish to indicate to users that they have to provide a GMT date, he can use two parameters: the first will contain the date itself and the second (e.g. {\it dateType}) will specify the nature of the first parameter. The service provider may then use all the richness of the PDL grammar for expressing conditions on/between these two parameters.

\section{The ParameterGroup Class}\label{par-group}

\begin{figure}[htbp]
\begin{center}
\includegraphics[width=0.7\textwidth]{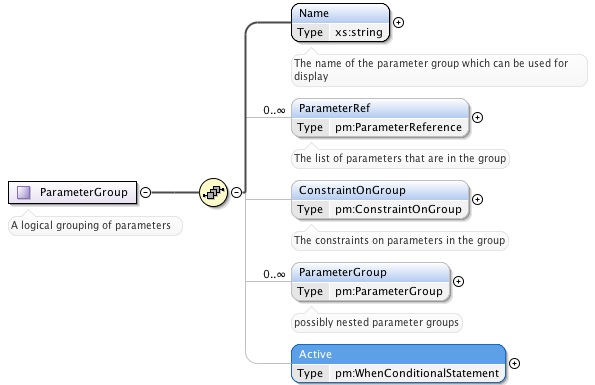} 
\caption{Graphical representation of the ParameterGroup Class}
\label{Pic-ParameterGroup}
\end{center}
\end{figure}

The {\it ParameterGroup} Class (see figure \ref{Pic-ParameterGroup}) is used for grouping
parameters according to a criterion of relevancy arbitrarily chosen by service provider (for instance
parameters may be grouped according to the physics : position-group, speed-group; thermodynamic-group).
However,  the ParameterGroup is not only a kind of parameter set, but also can be used for
defining complex relations and/or constraints involving the contained parameters (cf. paragraph
\ref{par-ConstraintsOnGroup}).\\
This Class {\bf must contain} a single Name. This name is a String and is the identification label
of the ParameterGroup, and two groups cannot have the same Name in a given PDL description instance.\\
Optional fields are
\begin{itemize}
\item the references to the parameters (cf. paragraph \ref{par-parRef}) one want to include into the
group;
\item a {\it ConstraintOnGroup} object (cf. paragraph 
\ref{par-ConstraintsOnGroup}). This object is used for expressing the complex relations and
constraints involving parameters.
\item an  {\it Active} field of type {\it WhenConditionalStatement} (cf. paragraph \ref{par-WhenConditionalStatment}). If there is no {\it Active} element the group will be considered active by default  (e.g. in case of a graphical representation it will be displayed). Otherwise, the activations depends on the result of the evaluation of the Criterion contained into the  When conditional statement (cf. paragraphs \ref{par-WhenConditionalStatment} and \ref{par-EvalCriteria}).
\item the {\it ParameterGroup} object contained within the current root group. Indeed the
{\it ParametersGroup} is a recursive object which can contain other sub-groups.
\end{itemize}

{\bf NB:} The name of every {\it ParameterGroup} is unique.\\

%% Agree with Paul view: at the beginning I (CMZ) introduced this constraint for avoiding contradictions.
%But one could also build contradiction by putting different constraints on a given group...
%{\bf NB:} A given {\it SingleParameter} object could only belong to one {\it
%ParameterGroup}\footnote{As we will see in paragraph \ref{par-ConstraintsOnGroup}, constraints on
%parameters are defined at the level of the group. If a {\it SingleParameter} belongs only to one
%group, it will be easier to verify that there is no contradictions on conditions}.
{\bf NB:}  For any practical use, the number on the parameter referenced into a given group summed
to the number of sub-groups of the same group must be greater than one. Otherwise the group would be
a hollow shell.

\section{The Expression Class}\label{par02}
The {\it Expression} is the most versatile component of the PDL. It occurs almost everywhere: in
defining fields for {\it SingleParameters} (cf. paragraph \ref{par01}) or in defining conditions and
criteria).\\
Expression itself is an abstract Class. In this section we are going to review all the concrete
Class extending and specializing expressions.\\

 {\bf N.B.} In what follows, we will call a {\bf numerical expression} every {\it
expression} involving only numerical types. This means that the evaluation of such expressions
should lead to a number (or a vector number if the dimension of the expression is greater than
one).\\

\subsection{The AtomicParameterExpression}\label{par02_01}
\begin{figure}[htbp]
\begin{center}
\includegraphics[width=0.7\textwidth]{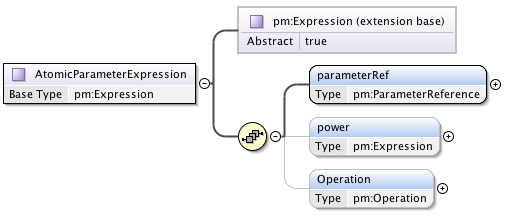} 
\caption{Graphical representation of the AtomicParameterExpression Class}
\label{Pic-AtomicParameter}
\end{center}
\end{figure}
The {\it AtomicParameterExpression} (extending {\it Expression}, see figure
\ref{Pic-AtomicParameter}) is the simplest expression that could be built involving a defined
parameter. This Class {\bf must contain} unique reference to a given parameter.\\

Optional fields, valid only for numerical types, are :
\begin{itemize}
\item A  {\bf numerical} power {\it Expression} object;
\item An  {\it Operation} object (cf. paragraph \ref{par02_02}).\\
\end{itemize}
Let $p$ and $exp$ be respectively the parameter and the power expression we want to encapsulate. The
composite object could be presented as follows:
\begin{equation}\label{eq01}
 \underbrace{  p^{exp} \underbrace{  \overbrace{\left( \begin{array}{c} + \\ - \\ \ast  \\ \cdot \\ \div   \end{array} \right) }^{\mbox{\tiny Operation type}}
 \overbrace{    \left( \mbox{AnotherExpression}\right) }^{\mbox{\tiny expression contained in operation}}   }_{\mbox{\tiny Operation object}}}_{\mbox{\tiny Atomic Parameter Expression}}
\end{equation}

To evaluate a given {\it AtomicParameterExpression}, one proceeds as follows: 
Let $d_p$, $d_{exp}$ be respectively the dimension of the parameter $p$ referenced, the
dimension of the power expression and the dimension of the expression contained within the operation
object.\\
The exponent part of the expression is legal if and only if:
\begin{itemize}
\item $d_p=d_{exp}$. In this case $p^{exp}$ is a $d_p$-size vector expression and $\forall$
$i=1,...,d_p$ the $i$ component of this vector is equal to ${p_i}^{exp_i}$, where $p_i$ is the value
of the $i$ component of vector parameter $p$ and $exp_i$ is the value obtained by interpreting the
$i$ component of vector expression $exp$.
\item Or $d_{exp}=1$. In this case, $\forall$ $i=1,...,d_p$ the $i$ component of the vector result
is equal to ${p_i}^{exp}$, where $p_i$ is the same as defined above.\\
\end{itemize} 

Whatever the method used, let us note $ep$ the result of this first step. We recall that the
dimension of $ep$ is always equal to $d_p$. In order to complete the evaluation of the expression,
one should proceed as shown in paragraph \ref{par02_02}, by setting there $b=ep$.

\subsection{The AtomicConstantExpression}\label{par02_03}
\begin{figure}[htbp]
\begin{center}
\includegraphics[width=0.7\textwidth]{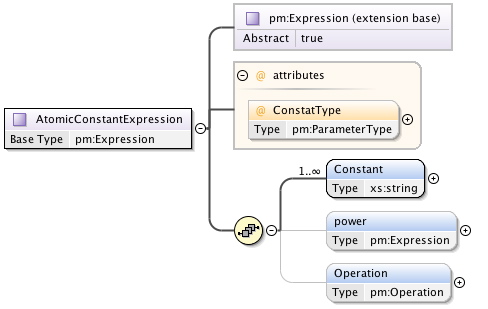} 
\caption{Graphical representation of the AtomicConstantExpression Class}
\label{Pic-AtomicConstant}
\end{center}
\end{figure}

The {\it AtomicConstantExpression} (extending {\it Expression}, see figure \ref{Pic-AtomicConstant})
is the simplest expression that could be built involving constants. Since this class could be used
for defining a constant vector expression, it {\bf must contain}
\begin{itemize}
\item A single list of String which expresses the value of each component of the expression. Let
$d_c$ be the size of the String list. If $d_c=1$ the expression is scalar and it is a vector
expression if $d_c>1$.
\item An attribute {\it ConstantType} of type {\it ParameterType} (cf. paragraph
\ref{par-ParameterType}) which defines the nature of the constant expression. The allowed
types are the same as in the field {\it parameterType} of the Class {\it SingleParameter}.
\end{itemize}
The Class {\bf is legal if and only if} every element of the String list could be cast into the
type expressed by the attribute {\it constantType}.\\

Optional fields, valid only for numerical types, are : 
\begin{itemize}
\item A  {\bf numerical} power {\it Expression} object;
\item An {\it operation} object (cf. paragraph \ref{par02_02}).
\end{itemize}

Let $s_i$ ($i=1,...,d_c$) and $exp$ be respectively the $i$ component of the String list and the
power expression we want to encapsulate. The composite Class could be presented as follows:
\begin{equation}
 \underbrace{    \overbrace{ \left( s_1  , s_2, ..., s_{d_c}\right)^{exp} }^{\mbox{\tiny List of String  to cast into the provided type}}   \underbrace{  \overbrace{\left( \begin{array}{c} + \\ - \\ \ast  \\ \cdot \\ \div   \end{array} \right) }^{\mbox{\tiny Operation type}}
 \overbrace{    \left( \mbox{AnotherExpression}\right) }^{\mbox{\tiny expression contained in operation}}   }_{\mbox{\tiny Operation object}}}_{\mbox{\tiny Atomic Constant Expression}}
\end{equation}
To evaluate a given {\it AtomicConstantExpression}, one proceeds as follows: let $d_c$, $d_{exp}$
be respectively the dimension of the vector constant ($d_c$ is equal to one in case of scalar constant),
the dimension of the power expression and the dimension of the expression contained within the operation object.\\
The exponent part of the expression is legal if and only if:
\begin{itemize}
\item $d_c = d_{exp}$. In this case $(s_1,...,s_{d_c})^{exp}$ is a $d_c$ size vector expression and
$\forall i =1,...,d_c$ the $i$-th component of this vector is equal to $s_i^{exp_i}$, where $exp_i$
is the value obtained by interpreting the $i$ component of vector $exp$.
\item Or $d_{exp}=1$. In this case, $\forall i =1,...,d_c$ the $i$ component of the vector result is
equal to $s_i^{exp}$.\\
\end{itemize}
Whatever the method used, let us note $ep$ (whose dimension is always equal to $d_c$)  is the result
of this first step.  In order to complete the evaluation of the expression, one should proceed as
exposed in paragraph \ref{par02_02}, by substituting there $b=ep$.

\subsection{The ParenthesisContentExpression Class}\label{par02_04}

\begin{figure}[htbp]
\begin{center}
\includegraphics[width=0.7\textwidth]{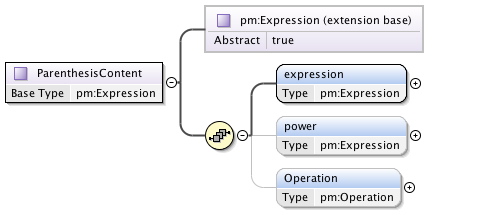} 
\caption{Graphical representation of the ParenthesisContent expression Class}
\label{Pic-ParenthesisContent}
\end{center}
\end{figure}

The {\it parenthesisContent} Class (extending {\it Expression}, see
\ref{Pic-ParenthesisContent}) is used to explicitly denote precedence by grouping the
expressions that should be evaluated first. This Class {\bf must contain} a single {\bf numerical}
object $Expression$ (referred to hereafter as $exp_1$).\\
Optional fields are 
\begin{itemize}
\item A {\bf numerical} power {\it expression} object (referred to hereafter as $exp_2$);
\item An {\it Operation} object (cf. paragraph \ref{par02_02}).\\
\end{itemize}
This composite Class could be presented as follows:
\begin{equation}
 \underbrace{    \underbrace{ \left( exp_1\right) }_{\mbox{\tiny Priority term}} ^{\hspace{15mm}exp_2} \underbrace{  \overbrace{\left( \begin{array}{c} + \\ - \\ \ast  \\ \cdot \\ \div   \end{array} \right) }^{\mbox{\tiny Operation type}}
 \overbrace{    \left( \mbox{AnotherExpression}\right) }^{\mbox{\tiny expression contained in operation}}   }_{\mbox{\tiny Operation object}}}_{\mbox{\tiny Parenthesis Expression}}
\end{equation}
In order to evaluate this object expression, one proceeds as follows: first one evaluates the
expression $exp_1$ that has the main priority. Then one proceeds exactly as in paragraph
\ref{par02_01} (after the equation (\ref{eq01})) by substituting $p=exp_1$ and $exp=exp_2$.

\subsection{The Operation Class}\label{par02_02}
\begin{figure}[htbp]
\begin{center}
\includegraphics[width=0.6\textwidth]{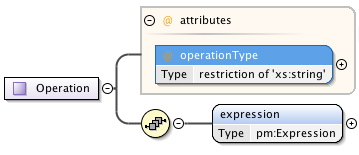} 
\caption{Graphical representation of Operation Class}
\label{Pic-Operation}
\end{center}
\end{figure}

The {\it Operation} Class (see figure \ref{Pic-Operation}) is used for expressing operations
involving two {\bf numerical} expressions. This Class {\bf must contain}:
\begin{itemize}
\item an {\it operationType} attribute. This attribute could take the following values: plus for the
sum, minus for the difference, multiply for the standard product, scalarProduct for the scalar
product and divide for the standard division. Hereafter these operators will be respectively denoted
$+,-,\ast,\cdot, \div$.
\item an {\it Expression} Class.
\end{itemize}
\begin{equation}\label{OperationEquation}
\underbrace{ \overbrace{\left( \begin{array}{c} + \\ - \\ \ast  \\ \cdot \\ \div   \end{array} \right) }^{\mbox{\tiny Operation type}}
 \overbrace{    \left( \mbox{ContaindedExpression}\right) }^{\mbox{\tiny expression contained in operation}}   }_{\mbox{\tiny Operation object}}
\end{equation}

The {\it Operation} Class is always contained within a {\bf numerical} {\it Expression} (cf.
paragraph \ref{par02}) and could not exist alone.
Let $a$ be the result of the evaluation of the expression object containing the
operation\footnote{this came from the evaluation of parameterRef field in case of an {\it
AtomicParameterExpression}  (cf. paragraph \ref{par02}), from the evaluation of constant field in the
case of a {\it AtomicConstantExpression} (cf. paragraph \ref{par02_03}), from the evaluation of the Expression field in case of an  {\it parenthesisContentExpression}  (cf. paragraph \ref{par02_04}) and from the evaluation of the Function object in case of a {\it FunctionExpression} (cf. par. \ref{functionExpressionPar}) }  let $b$ the
result of the evaluation of the {\bf numerical} expression contained within the operation. As usual,
we note $d_a$ and $d_b$ the dimensions of $a$ and $b$.\\

The operation evaluation is legal if and only if:
\begin{itemize}
\item $d_a=d_b$ and operation type (i.e. the operator) $op \in \{ + , - , \ast , \div \}$. In this
case $a \, op \, b$ is a vector expression of size $d_a$ and $\forall$ $i=1,...,d_a$ the $i$
component of this vector is equal to $(a_i \, op \,b_i)$ (i.e. a term by term operation).
\item Or $d_a = d_b$ and operation type $op$ is `$\cdot$'. In this case $a \cdot b$ is the result of
the scalar product $\sum_{i =1}^{d_a} a_i \ast b_i$. It is obvious that the dimension of this result
is equal to $1$.
\item Or $d_b=1$ and operation type (i.e. the operator) $op \in \{ + , - , \ast , \div \}$. In this
case $a \, op \, b$ is a vector expression of size $d_a$ and $\forall$ $i=1,...,d_a$ the $i$
component of this vector is equal to $(a_i \, op \,b)$.
\item Or $d_a=1$ and operation type (i.e. the operator) $op \in \{ + , - , \ast , \div \}$. This
case in symmetric to the previous one.
\end{itemize}

The type of the result is automatically induced by a standard cast operation performed during the
evaluations (for example a double vector added to an integer vector is a double vector).

\subsection{The FunctionType Class}\label{par-FunctionType}
This Class is used for specifying the mathematical nature of the function contained within a {\it
Function} object (cf. paragraph \ref{par02_05}). The unique String field this Class contains could
take one of these values:
size, abs, sin, cos, tan, asin, acos, atan, exp, log, sum, product. In paragraph \ref{par02_05} it
is explained how these different function types are used and handled.

\subsection{The Function Class}\label{par02_05}
\begin{figure}[htbp]
\begin{center}
\includegraphics[width=0.6\textwidth]{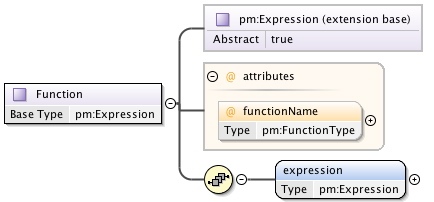} 
\caption{Graphical representation of Function Class}
\label{Pic-Function}
\end{center}
\end{figure}
The {\it Function} Class (extending {\it expression}, see figure \ref{Pic-Function}) is used for
expressing a mathematical function on expressions.
This Class {\bf must contain} 
\begin{itemize}
\item A {\it functionName} attribute (of type {\it functionType} (cf. paragraph
\ref{par-FunctionType})) which specifies the nature of the function.
\item An {\it Expression} object (which is the function argument).\\
\end{itemize}
Let $arg$ be the result of the evaluation of the function argument expression and  $d_{arg}$ its
dimension.
The {\it function} Class evaluation {\bf is legal if and only if}:
\begin{itemize}
\item $f  \in \{ \mbox{abs, sin, cos, tan, asin, acos, atan, exp, log} \}$ and the function argument
is a {\bf numerical} expression. In this case the  result  is a $d_{arg}$-size vector and each
component  $r_i = f(arg_i$), $\forall \, i=1,...,d_{arg}$.
\item Or $f=$sum (likewise $f=$product) and the argument is a {\bf numerical} expression. In this
case the result is a scalar value equal to $\sum_{i=1}^{i=d_{arg}} arg_i$ (likewise $\prod
_{i=1}^{i=d_{arg}} arg_i$), where $arg_i$ is the value obtained
by interpreting the $i$ component of vector expression $arg$.
\item Or $f=$size. In this case the result is the scalar integer value $d_{arg}$. 
\end{itemize}
From what we saw above, the result of the interpretation of a function Class { \bf is always a number}.

\subsection{The FunctionExpression Class} \label{functionExpressionPar}
\begin{figure}[htbp]
\begin{center}
\includegraphics[width=0.6\textwidth]{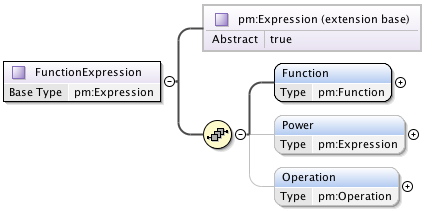} 
\caption{Graphical representation of FunctionExpression Class}
\label{Pic-FunctionExpression}
\end{center}
\end{figure}
The {\it FunctionExpression} Class (extending {\it Expression}, see figure
\ref{Pic-FunctionExpression}) is used for building mathematical expressions involving functions.\\
This Class {\bf must contains} a single {\it Function} object (cf. paragraph \ref{par02_05}).\\
Optional fields, valid only for numerical types, are :
\begin{itemize}
\item A {\bf numerical} power {\it Expression} object;
\item An {\it Operation} object� (cf. paragraph \ref{par02_02}).\\
\end{itemize}

This composite Class could be presented as follows:
\begin{equation}
 \underbrace{    \underbrace{ \left(\mbox{function}\right) }_{\mbox{\tiny Function object}} ^{\hspace{15mm}exp} \underbrace{  \overbrace{\left( \begin{array}{c} + \\ - \\ \ast  \\ \cdot \\ \div   \end{array} \right) }^{\mbox{\tiny Operation type}}
 \overbrace{    \left( \mbox{AnotherExpression}\right) }^{\mbox{\tiny expression contained in operation}}   }_{\mbox{\tiny Operation object}}}_{\mbox{\tiny FunctionExpression Object}}
\end{equation}
In order to evaluate this Class expression, one proceed as follows: first one evaluate the funtion
expression as explained in paragraph  \ref{par02_05}. Then one proceed exactly as in paragraph
\ref{par02_01} (after the equation (\ref{eq01})) by taking $p=$function.

\section{Expressing complex relations and constraints on parameters}\label{complexRelations}
In this part of the document we will explain how PDL objects could be used for building complex
constraints and conditions involving input and/or output parameters.

\subsection{The ConstraintOnGroup Object}\label{par-ConstraintsOnGroup}
\begin{figure}[htbp]
\begin{center}
\includegraphics[width=0.6\textwidth]{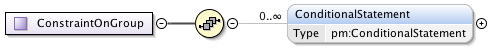} 
\caption{Graphical representation of ConstraintOnGroup object}
\label{Pic-ConstraintOnGroup}
\end{center}
\end{figure}

The {\it  ConstraintOnGroup} object (see figure \ref{Pic-ConstraintOnGroup}) is always contained within a {\it ParameterGroup} object and could not exist alone.
This object {\bf must contain} the {\it ConditionalStatement} objects. The latter are used, as is
shown in paragraph \ref{par-ConditionalStatement}, for expressing the complex relations and
constraints involving parameters.

\subsection{The ConditionalStatement object}\label{par-ConditionalStatement}
The {\it ConditionalStatement} object is abstract and, as its name suggests, is used for defining conditional statements. In this section we are going to review the two concrete objects
extending and specializing  {\it ConditionalStatement}.

\subsubsection{The AlwaysConditionalStatement}\label{par-AlwaysConditionalStatement}
\begin{figure}[htbp]
\begin{center}
\includegraphics[width=1.1\textwidth]{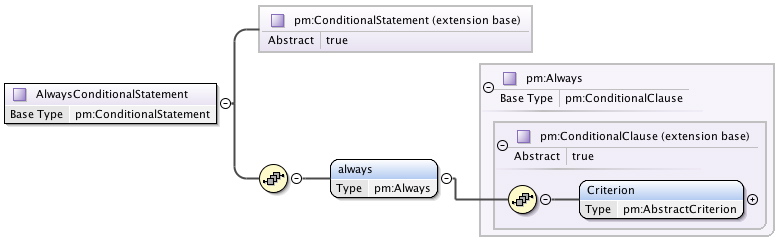} 
\caption{Graphical representation of AlwaysConditionalStatement object}
\label{Pic-AlwaysConditionalStatement}
\end{center}
\end{figure}
This object (see figure \ref{Pic-AlwaysConditionalStatement}), as it name suggests,  is used
for expressing statement that must always be valid. It {\bf must contain} a single {\it Always}
object (which extends {\it ConditionalClause}, cf. paragraph \ref{par-ConditionalClause}).

\subsubsection{The IfThenConditionalStatement}\label{par-IfThenConditionalStatement}
\begin{figure}[htbp]
\begin{center}
\includegraphics[width=1.0\textwidth]{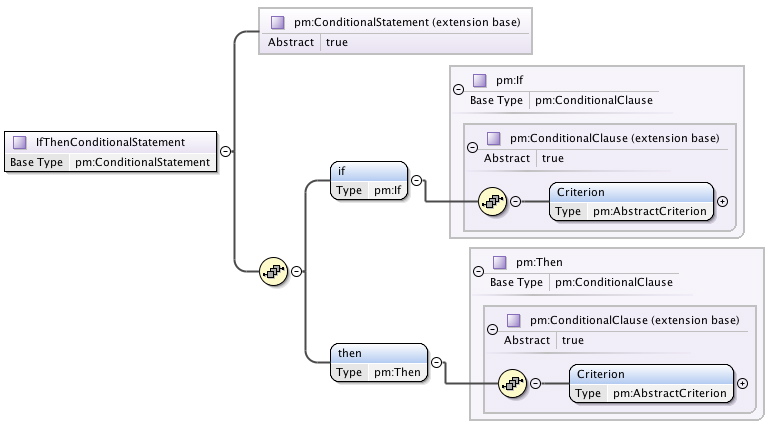} 
\caption{Graphical representation of IfThenConditionalStatement object}
\label{Pic-IfThenConditionalStatement}
\end{center}
\end{figure}
This object (see figure \ref{Pic-IfThenConditionalStatement}), as it name suggests, is used for
expressing statements that are valid only if a previous condition is verified. It {\bf must
contain}:
\begin{itemize}
\item an  {\it If} object (which extends {\it ConditionalClause}, cf. paragraph
\ref{par-ConditionalClause}).
\item a {\it Then} object (which extends {\it ConditionalClause}, cf. paragraph
\ref{par-ConditionalClause}).
\end{itemize}
If the condition contained within the {\it If} object is valid, the condition contained within
the {\it Then} object {\bf must be} valid too.

\subsubsection{The WhenConditionalStatement object}\label{par-WhenConditionalStatment}
The when conditional statement is valid when the enclosed {\it When} conditional clause
evaluates to true (cf. paragraph \ref{par-EvalCriteria}). It contains a unique field of {\it When} type (cf. paragraph \ref{par-ConditionalClause}). It was introduced for the purpose of dealing with the case of activating a ParameterGroup (cf paragraph \ref{par-group}): Thus When has the advantage of having a restricted form of ConditionalStatement that could have no {\it side effects} in the Then part.
\begin{figure}[htbp]
\begin{center}
\includegraphics[width=1.0\textwidth]{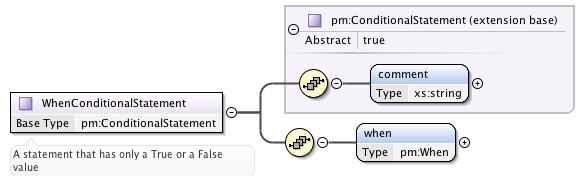} 
\caption{Graphical representation of a WhenConditionalStatement object}
\label{Pic-WhenConditionalStatement}
\end{center}
\end{figure}

\subsection{The ConditionalClause object}\label{par-ConditionalClause}
\begin{figure}[htbp]
\begin{center}
\includegraphics[width=0.75\textwidth]{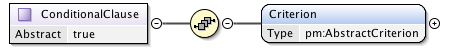} 
\caption{Graphical representation of ConditionalClause object}
\label{Pic-ConditionalClause}
\end{center}
\end{figure}
\begin{figure}[htbp]
\begin{center}
\includegraphics[width=0.7\textwidth]{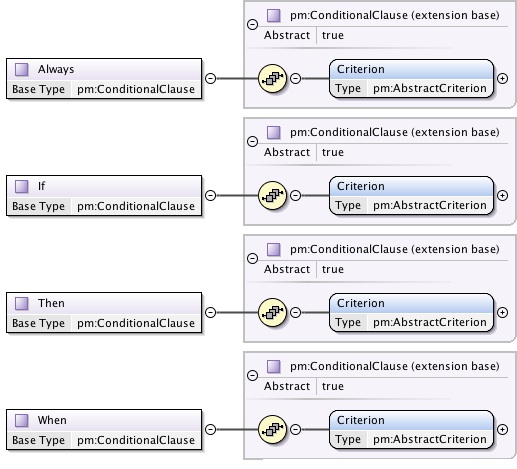} 
\caption{Graphical representation of Always, If, Then and When clauses}
\label{Pic-ConcreteClause}
\end{center}
\end{figure}

The {\it ConditionalClause} object (see figure \ref{Pic-ConditionalClause}) is abstract. It {\bf
must contain} a single {\it Criterion} object of type {\it AbstractCriterion} (cf. paragraph
\ref{par-AbstractCriterion}).\\
The four concrete objects extending the abstract {\it ConditionalClause} are (see figure
\ref{Pic-ConcreteClause}):
\begin{itemize}
\item {\it Always};
\item {\it If};
\item {\it Then};
\item {\it When}.
\end{itemize}
The Criterion contained within a {\it Always} object must always evaluates to {\it true} (we will hereafter say it is valid, cf paragraph
\ref{par-AlwaysConditionalStatement}). With other words, this means that {\it it is good} only when the evaluation of the criterion contained into the {\it Always} object' evaluates to {\it true}. What if it is not good? It is wrong. {\it Wrong} evaluation is typically cached for notifying errors to users.\\

The Criterion contained within a {\it When} object will be valid only when the enclosed Expression
evaluates to True (cf. paragraphs \ref{par-WhenConditionalStatment} and \ref{par-EvalCriteria} ).

The {\it If} and {\it Then} objects work as a tuple by composing the  {\it
IfThenConditionalStatement} (cf. paragraph
\ref{par-IfThenConditionalStatement}).

\subsection{The AbstractCriterion object}\label{par-AbstractCriterion}
\begin{figure}[htbp]
\begin{center}
\includegraphics[width=0.7\textwidth]{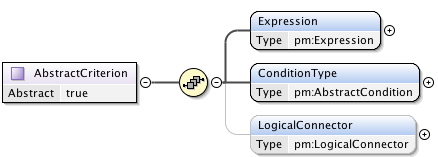} 
\caption{Graphical representation of AbstractCriterion object}
\label{Pic-AbstractCriterion}
\end{center}
\end{figure}
The objects extending {\it AbstractCriterion} (see figure \ref{Pic-AbstractCriterion}) are
essentials for building {\it ConditionalStatemets} (cf. paragraph \ref{par-ConditionalStatement})
since they are contained within the {\it Always, If} and {\it Then} objects (cf. paragraph
\ref{par-ConditionalClause}).
An {\it AbstractCriterion} object {\bf must contain}:
\begin{itemize}
\item an {\it Expression} object (cf. paragraph \ref{par02});
\item a {\it ConditionType} which is an object of type {\it AbstractCondition} (cf. paragraph 
\ref{par-ConditionType}).
This object specify which condition must be satisfied by the previous {\it Expression}.
\end{itemize}
An optional field is the single {\it LogicalConnector} object (cf. paragraph
\ref{par-LogicalConnector}) used for building logical expressions.\\
The two concrete objects extending  {\it AbstractCriterion}  are {\it Criterion} and {\it
ParenthesisCriterion}.  The latter of these two objects allows for assigning priority when
interpreting and linking the criteria (cf. paragraph \ref{par-EvalCriteria}).

\subsubsection{The Criterion object}
\begin{figure}[htbp]
\begin{center}
\includegraphics[width=0.7\textwidth]{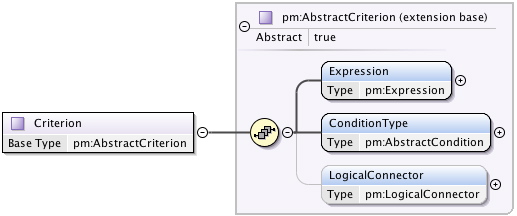} 
\caption{Graphical representation of Criterion object}
\label{Pic-Criterion}
\end{center}
\end{figure}
This object (see figure \ref{Pic-Criterion}) extends the  {\it AbstractCriterion} without
specializing it. It is indeed just a concrete version of the abstract type.

\subsubsection{The ParenthesisCriterion object}
\begin{figure}[htbp]
\begin{center}
\includegraphics[width=0.8\textwidth]{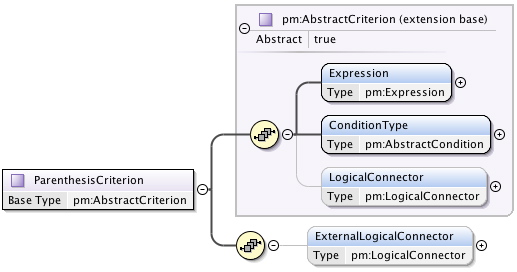} 
\caption{Graphical representation of ParenthesisCriterion object}
\label{Pic-ParenthesisCriterion}
\end{center}
\end{figure}
This object (see figure \ref{Pic-ParenthesisCriterion}) extends and specialize the  {\it
AbstractCriterion}. It is used for defining arbitrary priority in interpreting boolean expression
based on criteria.
The optional field of {\it ParenthesisCriterion} is an {\it ExternalLogicalConnector} object
of type {\it LogicalConnector}. It is used for linking other criteria, out of the priority
perimeter defined by the parenthesis (cf.  paragraph \ref{par-EvalCriteria}).

\subsection{The LogicalConnector object}\label{par-LogicalConnector}
\begin{figure}[htbp]
\begin{center}
\includegraphics[width=0.7\textwidth]{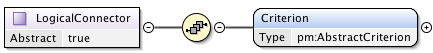} 
\caption{Graphical representation of LogicalConnector object}
\label{Pic-LogicalConnector}
\end{center}
\end{figure}
The {\it LogicalConnector} object (see figure \ref{Pic-LogicalConnector}) is used for building
complex logical expressions. It is an abstract object and it {\bf must} contain a Criterion
of type {\it AbstractCriterion} (cf. paragraph \ref{par-AbstractCriterion}).\\
The two concrete objects extending {\it LogicalConnector} are:
\begin{itemize}
\item the {\it And} object used for introducing the logical AND operator between two
criteria;\footnote{The first criterion is the one containing the {\it LogicalConnector} and the
second is the criterion contained within the connector itself.}
\item the {\it Or} object used for introducing the logical OR operator between two criteria.
\end{itemize}

\subsection{The AbstractCondition object}\label{par-ConditionType}
{\it AbstractCondition} is an abstract object. The objects extending it always belong to an  {\it 
AbstractCriterion} (cf. \ref{par-AbstractCriterion}). In this context, they are used combined with
an {\it Expression} object, for expressing the condition that the expression must satisfy.\\
Let us consider a given criterion object $\mathcal{CR}$ (extending{\it AbstractCriterion})  and let
us note $\mathcal E$ and $\mathcal C$ the expression and the condition contained within
$\mathcal{CR}$.
In what follows we are going to explain the different objects specializing  {\it AbstractCondition}
and their behavior.

\subsubsection{The IsNull condition}\label{par-IsNull}
This object is used for specifying that the expression $\mathcal E$ has no assigned value (this is
exactly the same concept as the NULL value in Java or the None value in Python).
Indeed, if and only if $\mathcal E$ has no assigned value, the evaluation of the tuple $(\mathcal
E, \mathcal C)$ leads to a TRUE boolean value. Thus, in the case $\mathcal{CR}$ has no {\it
LogicalConnector}, the criterion is true.

\subsubsection{The `numerical-type' conditions}
These objects are used for specifying that the result of the evaluation of the expression $\mathcal
E$ is of a given numerical type. The tuple $(\mathcal E, \mathcal C)$ is legal if and only if
$\mathcal E$ is a {\bf numerical} expression. \\
The `numerical-type' objects extending {\it AbstractCondition}  are:
\begin{itemize}
\item {\it IsInteger}, in this case the evaluation of the tuple $(\mathcal E, \mathcal C)$ leads to
a TRUE boolean value if and only if the evaluation of the numerical expression $\mathcal E$ is an
integer.
\item {\it IsReal}, in this case the evaluation of the tuple $(\mathcal E, \mathcal C)$ leads to a
TRUE boolean value
if and only if the evaluation of the numerical expression $\mathcal E$ is a real number.
\end{itemize} 

\subsubsection{The BelongToSet condition}\label{par-BelongToSet}
\begin{figure}[htbp]
\begin{center}
\includegraphics[width=0.8\textwidth]{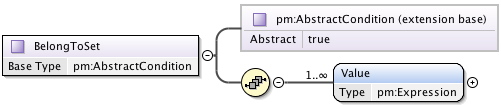} 
\caption{Graphical representation of BelongToSet object}
\label{Pic-BelongToSet}
\end{center}
\end{figure}
This object (see figure \ref{Pic-BelongToSet}) is used for specifying that the expression $\mathcal
E$ could take only a finite set of values.
It {\bf must contain} the {\it Values} (which are objects of type {\it Expression}) defining the set
of legal values. The number of {\it Values} must be greater than one.\\
This object is legal only if all the {\it Expressions} of the set are of the same type (e.g. they
are all numerical, or all boolean or all String expressions).\\
The tuple $(\mathcal E, \mathcal C)$ leads to a TRUE boolean value if and only if:
\begin{itemize}
\item the expression $\mathcal E$ and the expressions composing the set are of the same type
\item and an element $\mathcal E_s$ exists in the set such that $\mathcal E_s = \mathcal E$.\\
This last equality is to be understood in the following sense: let $=_t$ be the equality operator
induced by the type (for numerical type the equality is in the real number sense, for String type
the equality is case sensitive and for boolean the equality is in the classic boolean sense).\\
Two expressions are equal if and only if
\begin{itemize}
\item the expressions have the same size $d_{\mathcal E}$,
\item and $\mathcal E_s^i =_t \mathcal E^i$, $\forall i =1,...,d_{\mathcal E}$, where $\mathcal
E_s^i$ and $ \mathcal E^i$ are respectively the result of the evaluation of the
$i$ component of expressions $\mathcal E_s$ and $\mathcal E$. 
\end{itemize}
\end{itemize}

\subsubsection{The ValueLargerThan object}
\begin{figure}[htbp]
\begin{center}
\includegraphics[width=0.8\textwidth]{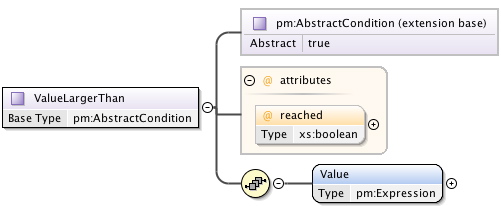} 
\caption{Graphical representation of ValueLargerThan object}
\label{Pic-ValueLargerThan}
\end{center}
\end{figure}
This object (see figure \ref{Pic-ValueLargerThan}) is used for expressing that the result of the
evaluation of the expression $\mathcal E$ must be greater than a given value.\\
It {\bf must contain}
\begin{itemize}
\item a {\bf numerical} {\it Expression} $\mathcal E_c$. 
\item a {\it Reached} attribute, which is a boolean type.
\end{itemize}
The tuple $(\mathcal E, \mathcal C)$ is legal only if $\mathcal E$ is a numerical expression.\\
This tuple leads to a TRUE boolean value if and only if the result of the evaluation of the
expression $\mathcal E$ is greater than the result of the evaluation
of the expression $\mathcal E_c$ and the attribute {\it Reached} is false. Otherwise if the {\it
Reached} attribute is true the expression $\mathcal E$ may be greater than or equal to the result.

\subsubsection{The ValueSmallerThan object}
\begin{figure}[htbp]
\begin{center}
\includegraphics[width=0.8\textwidth]{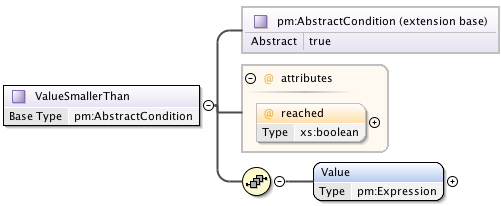} 
\caption{Graphical representation of ValueSmallerThan object}
\label{Pic-ValueSmallerThan}
\end{center}
\end{figure}
This object (see figure \ref{Pic-ValueSmallerThan}) is used for expressing that the result of the
evaluation of the expression $\mathcal E$ must be smaller than a given value.\\
It {\bf must contain}
\begin{itemize}
\item a  {\bf numerical} {\it Expression} $\mathcal E_c$. 
\item a  {\it Reached} attribute which is a boolean type.
\end{itemize}
The tuple $(\mathcal E, \mathcal C)$ is legal only if $\mathcal E$ is a numerical expression.\\
This tuple leads to a TRUE boolean value if and only if the result of the evaluation of the
expression $\mathcal E$ is smaller (otherwise smaller or equal when the  attribute {\it Reached} is
true) than the result of the evaluation of the expression $\mathcal E_c$.

\subsubsection{The ValueInRange object}
\begin{figure}[htbp]
\begin{center}
\includegraphics[width=0.8\textwidth]{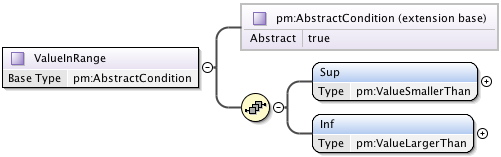} 
\caption{Graphical representation of ValueInRange object}
\label{Pic-ValueInRange}
\end{center}
\end{figure}
This object (see figure \ref{Pic-ValueInRange}) is used for expressing that the result of the
evaluation of the expression $\mathcal E$ must belong to a given interval. The definition of the
interval is made using the {\it ValueLargerThan} {\it ValueSmallerThan} objects.
Indeed, the {\it ValueInRange} object {\bf must contain}:
\begin{itemize}
\item an object  Inf of type {\it ValueLargerThan} for specifying the Inferior limit of the interval,
\item an object Sup of type {\it ValueSmallerThan} for specifying the Superior limit of the interval. 
\end{itemize}
The tuple $(\mathcal E, \mathcal C)$ is legal only if $\mathcal E$ is a numerical expression.\\
This tuple leads to a TRUE boolean value if and only if the evaluation of both tuples $(\mathcal E,
\mbox{\it ValueSmallerThan})$ and $(\mathcal E, \mbox{\it ValueLargerThan})$ lead to TRUE boolean
values.

\subsubsection{The ValueDifferentFrom object}
\begin{figure}[htbp]
\begin{center}
\includegraphics[width=0.8\textwidth]{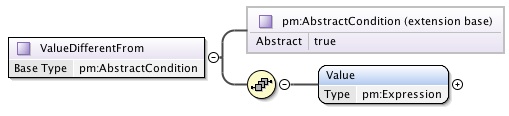} 
\caption{Graphical representation of ValueDifferentFrom object}
\label{Pic-ValueDifferentFrom}
\end{center}
\end{figure}
This object (see figure \ref{Pic-ValueDifferentFrom}) is used for specifying that the expression
$\mathcal E$ must be different from a given value.
It {\bf must contain} an {\it Expression} $\mathcal E_c$.\\
In order to be compared, the two expressions $\mathcal E$ and $\mathcal E_c$ must have the same
type.
The evaluation of the tuple $(\mathcal E, \mathcal C)$  leads to a TRUE boolean value only if
$\mathcal E \neq \mathcal E_c$. This inequality has to be understood in the sense explained in
paragraph \ref{par-BelongToSet} (in the second point of the list).

\subsubsection{The DefaultValue object}\label{par-DefaultValue}
\begin{figure}[htbp]
\begin{center}
\includegraphics[width=0.8\textwidth]{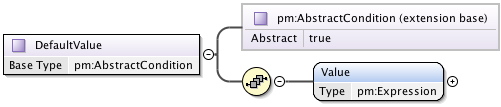} 
\caption{Graphical representation of DefaultValue object}
\label{Pic-DefaultValue}
\end{center}
\end{figure}
This object (see figure \ref{Pic-DefaultValue}) is used for specifying the default value of a
parameter.\\
It {\bf must contain} an {\it Expression} $\mathcal E_c$.\\
Since the default value of an expression involving functions, multiple parameters, etc. has no
particular sense, in the case of the present object the tuple $(\mathcal E, \mathcal C)$  is legal
only if
\begin{itemize}
\item $\mathcal E$ is an {\it AtomicParameterExpression} (cf. paragraph. \ref{par02_01}) \item and
the dimension and the type of the expression $\mathcal E_c$ are equal to the dimension and type
expressed in the {\it SingleParameter} object referenced into the {\it AtomicParameterExpression}.
\end{itemize}
Moreover, for having a legal {\it DefaultValue} object, the criterion $\mathcal{CR}$ containing it must be contained within 
the {\it Always} or {\it Then}  objects (cf. paragraph \ref{par-ConditionalClause}).

\subsection{Evaluating and interpreting criteria objects}\label{par-EvalCriteria}
The evaluation of the criterion type objects (cf. paragraph \ref{par-AbstractCriterion}) always
leads to a boolean value (the only exception is what we saw in paragraph \ref{par-DefaultValue},
where the criterion contains a {\it DefaultValue} condition).\\
We use hereafter the same notation introduced in \ref{par-ConditionType}: let us consider a given
criterion (extending{\it AbstractCriterion}) $\mathcal{CR}$ and let us note $\mathcal E$ and
$\mathcal C$ the expression and the condition contained within $\mathcal{CR}$.\\
When $\mathcal{CR}$ contains no {\it LogicalConnector} objects, the evaluation of the criterion is
straightforward :
the result is equal to the boolean-evaluation of the tuple $(\mathcal E, \mathcal C)$. This tuple
is evaluated according to the concrete class involved, as explained in paragraphs \ref{par-IsNull}
to \ref{par-DefaultValue} \\
It is a bit more complex when criteria contain {\it LogicalConnectors}. Let us see how to
proceed.\\
To begin with, let us consider only {\it Criterion} concrete objects:\\
As we saw in the previous paragraphs, criteria object are (with the help of {\it
LogicalConnectors} object) recursive and hierarchical objects.\\
This hierarchical structure composing a complex criterion could be graphically represented as
follows.
\begin{equation}\label{eq-CriterionStructure01}
(\mathcal E_1, \mathcal C_1) \xrightarrow[{\mbox{\tiny AND/OR}}]{LC_1} 
(\mathcal E_2, \mathcal C_2) \xrightarrow[{\mbox{\tiny AND/OR}}]{LC_2}
\cdots 
(\mathcal E_i, \mathcal C_i) \xrightarrow[{\mbox{\tiny AND/OR}}]{LC_i}
\cdots 
(\mathcal E_{N-1}, \mathcal C_{N-1}) \xrightarrow[{\mbox{\tiny AND/OR}}]{LC_{N-1}}
(\mathcal E_N, \mathcal C_N)
\end{equation}
where the index $1$, $i$ and $N$ are respectively for the root, the $i$ and the leaf criterion
composing the structure. The term $LC_i$ denotes the {\it LogicalConnector} contained within the
criterion $\mathcal{CR}_i$.\\
As we saw in paragraphs \ref{par-IsNull} to \ref{par-DefaultValue} every tuple $(\mathcal E_i,
\mathcal C_i)$, $i=1,..,N$ could be evaluated (according to the concrete object involved) and leads
to a boolean value $\mathcal B_i$. Thus the expression (\ref{eq-CriterionStructure01}) become
\begin{equation}\label{eq-CriterionStructure02}
\mathcal B_1 \xrightarrow[{\mbox{\tiny AND/OR}}]{LC_1} 
\mathcal B_2 \xrightarrow[{\mbox{\tiny AND/OR}}]{LC_2}
\cdots 
\mathcal B_i \xrightarrow[{\mbox{\tiny AND/OR}}]{LC_i}
\cdots 
\mathcal B_{N-1} \xrightarrow[{\mbox{\tiny AND/OR}}]{LC_{N-1}}
\mathcal B_N
\end{equation}
This last is a classic sequential boolean expression. It is evaluated from left to right and the
operator AND takes precedence over the OR operator.\\

Let us now consider {\it ParenthesisCriterion}  criteria. A representation of such a criterion
$\mathcal{CR}$ could be the following:
\begin{equation}
\Big \langle  (\mathcal E , \mathcal C) \xrightarrow{LC}  \mathcal{CR}_c  \Big \rangle_{\mathcal{CR}}  \xrightarrow{ELC}\,,
\end{equation}
where $\mathcal E$, $\mathcal C$, $LC$, $\mathcal{CR}_c$ are respectively 
the {\it Expression}, the condition, the {\it LogicalConnector} and the criterion contained within $LC$.
The term $ELC$ is the {\it ExternalLogicalConnector} of $\mathcal{CR}$.\\
The criterion structure contained within $\langle \cdot \rangle_{\mathcal{CR}}$ has the highest
priority and has to be evaluate, before the {\it ExternalLogicalConnector} evaluation.\\

In the case where $\mathcal{CR}_c$ is composed only of {\it Criterion} objects (so with no
{\it ParenthesisCriterion}), the evaluation of the content of  $\langle \cdot
\rangle_{\mathcal{CR}}$  is performed as shown before in (\ref{eq-CriterionStructure01}) and
(\ref{eq-CriterionStructure02}).

In the case where $\mathcal{CR}_c$ contains at least one {\it ParenthesisCriterion}, one has to go
deeper in the criterion structure to find the deepest criterion $\mathcal{CR}_d$ such that $\langle
\cdot \rangle_{\mathcal{CR}_d}$ contains only criteria of type {\it Criterion}.Thus one can simply
evaluate the content of  $\langle \cdot \rangle_{\mathcal{CR}_d}$ as already shown.\\

For illustrating how to proceed, let us consider the following complex-criterion structure:
\begin{equation}\label{eq-CriterionStructure03}
\begin{array}{l}
\displaystyle \Big \langle (\mathcal E_1, \mathcal C_1) \xrightarrow{LC_1} 
(\mathcal E_2, \mathcal C_2) \Big \rangle_{\mathcal{CR}_1 } \xrightarrow{ELC_1} 
\cdots \vspace{1mm}
\\
\hspace{1.5cm} \displaystyle \Big \langle (\mathcal E_{i-1}, \mathcal C_{i-1}) \xrightarrow{LC_{i-1}}
\Big \langle
(\mathcal E_i, \mathcal C_i) \xrightarrow{LC_i}
(\mathcal E_{i+1}, \mathcal C_{i+1}) \Big\rangle_{\mathcal{CR}_{i} }  \Big \rangle_{\mathcal{CR}_{i-1}}  \xrightarrow{ELC_{i-1}}  \vspace{1mm} \\
 \hspace{7cm} \cdots \displaystyle \Big \langle (\mathcal E_{N-1}, \mathcal C_{N-1}) \xrightarrow{LC_{N-1}}
(\mathcal E_N, \mathcal C_N) \Big\rangle_{\mathcal{CR}_{N-1} } \\
\end{array}
\end{equation}

From what we saw above, the expression (\ref{eq-CriterionStructure03}) becomes
\begin{equation}
\begin{array}{l}
\displaystyle \Big \langle \mathcal B_1 \xrightarrow{LC_1}  \mathcal B_2 \Big \rangle_{\mathcal{CR}_1} 
\xrightarrow{ELC_1}  \cdots \\
\displaystyle \hspace{3cm} \Big \langle \mathcal B_{i-1}  \xrightarrow{LC_{i-1}} \Big \langle \mathcal B_i 
 \xrightarrow{LC_i} \mathcal B_{i+1} \Big \rangle_{\mathcal{CR}_i} \Big \rangle_{\mathcal{CR}_{i-1}}  \xrightarrow{ELC_{i-1}}  \\
\displaystyle \hspace{7.5cm} \cdots  \Big \langle  \mathcal B_{N-1}   \xrightarrow{LC_{N-1}} \mathcal B_N  \Big \rangle_{\mathcal{CR}_{N-1}}
\end{array}
\end{equation}
and finally
\begin{equation}
\begin{array}{l}
\displaystyle \Big ( \mathcal B_1 \xrightarrow[{\mbox{\tiny AND/OR}}]{LC_1}  \mathcal B_2 \Big ) \xrightarrow[{\mbox{\tiny AND/OR}}]{ELC_1}  \cdots \\
\displaystyle \hspace{2.5cm} \Big ( \mathcal B_{i-1} \xrightarrow[{\mbox{\tiny AND/OR}}]{LC_{i-1}} \Big (
\mathcal B_i \xrightarrow[{\mbox{\tiny AND/OR}}]{LC_i} \mathcal B_{i+1} \Big) \Big )  \xrightarrow[{\mbox{\tiny AND/OR}}]{ELC_{i-1}}  \\
\displaystyle  \hspace{7.5cm}  \cdots \Big (   B_{N-1}   \xrightarrow[{\mbox{\tiny AND/OR}}]{LC_{N-1}} \mathcal B_N \Big ) \,. \\
\end{array}
\end{equation}
This last is a classical sequential boolean expression. It is evaluated from the left to the right.
The sub-expression between the parenthesis must be evaluated with the highest priority and the
operator AND takes precedence over the OR operator.

\section{PDL and formal logic}
We recall that PDL is a grammar and syntax framework for describing parameters and their
constraints. Since the description is rigorous and unambiguous, PDL could verify if the instance of
a given parameter is consistent with the provided description and related constraints. For example,
consider the description
\begin{equation}
\left\{
\begin{array}{l}
p_1 \mbox{is a Kelvin temperature}\\
\mbox{Always }  p_1 > 0 \\
\end{array}
\right..
\end{equation}
According to the description, the PDL framework could automatically verify the validity of the
parameter provided by the user.
If he/she provides $p_1=-3$, then this value will be rejected.\\
In any case PDL is not a formal-logic calculation tool. One could build the following description
with no problem:
\begin{equation}
\left\{
\begin{array}{l}
p_1 \in \mathbb R\\
\displaystyle \mbox{Always } \big(  (p_1 > 0) \mbox{ AND } (p_1 < 0) \big)\\
\end{array}
\right..
\end{equation}
The PDL language grammar is not a tool with capabilities to perceive logical contradictions which may be contained within statements. This kind of consideration is outside of the scope of the present standard.
For this reason, a validation system for PDL definitions is not required to implement the detection of contradictions, but may do if the services within its description realm make this feasible.
The current PDL reference implementation does not perform such contradiction detection and thus any parameter $p_1$ provided by user would be rejected for this example.\\
In other words {\bf people providing descriptions of services must pay great attention to their contents.}\\

\section{Remarks for software components implementing PDL}\label{SoftwareImplementation}
Throughout this document we have described PDL as a {\it grammar}. 
If we consider it just as a grammar, then a specific description should be considered as an implementation.\\ 
We remember that, since a PDL description is detailed, it is {a priori} possible to write once for all generic software components. These components will be automatically {\it configured} by a PDL description thus becoming {\it ad hoc} implementation software for the described service. Moreover checking algorithms could also be generated automatically starting from a description instance.
In our implementations we wanted to check practically that these concepts implied in the definition of PDL really works. 
The development of operational services (as the Paris-Durham shock code) also permits to ensure the coherence of the core grammar and to verify if the PDL's capabilities could meet the description needs of state of the art simulation codes.

At the present (Fall 2013) four software elements are implemented around PDL:
\begin{itemize}
\item the standalone dynamic client. It embeds the automatic generation of the verification layer (Google code repository at \url{https://code.google.com/p/vo-param/}). This development shows that a PDL description instance can be used for generating the checking algorithms and for generating a client with a dynamic-intelligent behavior helping the user in interacting with a service. This client could be used for interacting with services exposed using different job systems;
\item a server for exposing any exiting code as a web services. It embeds the verification layer. This development was done for showing that a PDL description instance can be used for generating the {\it ad hoc} server, exposing the described service. A particular feature of this server is that it can generates grids of model starting from a single job submission, which indicates ranges for parameters (GitHub repository at \url{https://github.com/cmzwolf});
\item the Taverna Plugin \cite{AstroTaverna}. From one point of view this plugin could be seen as an alternate client to the standalone one. From another point of view it is strongly oriented towards general physical and scientific interoperability (discussed in paragraph \ref{ParInteropIssues}) since it uses PDL description for validating the chaining of jobs composing a workflow. As the dynamic client, the Taverna plugin can be used for interacting with services exposed different job systems (GitHub repository for stable version at \url{https://github.com/wf4ever/astrotaverna}).  
\item the description editor, for editing PDL description from a Graphical User Interface. Since the key point for using PDL and take advantage of the software tools we have just described is a PDL description, we decided to provide the community with a tool for easily composing PDL description. In some sense this is the entry-point of the PDL software farmework  (google code repository at \url{https://code.google.com/p/pdl-editor/}).
\end{itemize}
All these developments validate the concepts of automatic generation of algorithms and the possibility of configuring, with highly specialized individual behaviors, generic software components. This is very important since it reduces drastically the development time for building services based on the PDL grammar. This is essential in a scientific context where only few scientists have access to software engineer for their IVOA developments.\\

 In further developments, PDL-client implementations will include a formal-logic module. This will permit finding contradictions inside the descriptions.\\
Such a  module will also be required for implementing the automatic computation of {\it a priori interoperability graphs}. 
It will also permit checking interoperability in terms of semantic annotations: for example, let A be the concept that describes an input parameter of a service $\mathcal S$ and B the concept that describes an output parameter of a service $\mathcal S'$. If A and B are the same concept, then both services match the interoperability criterion. However, if A and B are not the same concept we need, for piping the result of  
$\mathcal S'$ to $\mathcal S$, to ask if the concept B is more specific than the concept A, in other words, if the concept B is generalized or subsumed by the concept A. If this happens then both services match again the interoperability criterion. 
Interoperability only makes sense when there is an application or infrastructure that allows communication and connection of different services. An example is the applications for orchestrating services by designing workflows (as described in section 2.2). Further developments for PDL include the implementation of interoperability mechanisms in Taverna.

\section{Annex}

\subsection{A practice introduction to PDL (or dive in PDL)}\label{divePDL}
In this section we present a practice approach to PDL. It is inspired by one of the first services we deployed using the PDL framework: the Meudon Stark-H broadening computation service for Hydrogen (\href{http://atomsonline.obspm.fr}{http://atomsonline.obspm.fr}).\\
The exposed code take as input four parameters:
\begin{itemize}
\item A quantum number $N_i$, which corresponds to the upper energy level.
\item A quantum number $N_f$, which corresponds to the lower energy level.
\item A temperature $T$, which is the temperature of the simulated medium.
\item A density $\rho$, which is an electron density.
\end{itemize} 
With the existing exposure systems (mostly Soap \cite{Soap}, REST \cite{REST}, servlet web services) the information about parameters is more or less limited to a basic {\it function signature}: the two first parameters are {\it Integer}, while the two last are {\it Double}. But this information is not sufficient for a user wishing to use the service without knowing {a priori} the code: what are the unit of these parameters? What are their physical meaning?
PDL is a unified way for providing user with this information by hardcoding it directly in the software composing the service. With PDL service provider can easily express that 
\begin{itemize}
\item $N_i$ is {\it Integer}, it corresponds to the principal quantum number of the upper energy level and, as a quantum number, it has no dimension. The PDL {\it translation} of this sentence is:
\begin{lstlisting}[style=listXML]
<parameter dependency="required">
	<Name>InitialLevel</Name>
	<ParameterType>integer</ParameterType>
	<SkosConcept>http://example.edu/skos/initialLevel</SkosConcept>
	 <Unit>None</Unit>
	 <Dimension xsi:type="AtomicConstantExpression" ConstantType="integer">
		<Constant>1</Constant>
	</Dimension>
</parameter>
\end{lstlisting}
The PDL description points to the skos uri containing the definition of the physical concept. Moreover it says that the current parameter has $1$ as dimension. This means that the parameter is scalar (a dimensions greater than one is for vector parameters). The required attribute indicate that the user must submit this parameter to the service, and it is not optional.

\item $N_f$ is {\it Integer}, it corresponds to the principal quantum number of the lower energy level and, as a quantum number, it has no dimension. The PDL {\it translation} of this sentence is:
\begin{lstlisting}[style=listXML]
<parameter dependency="required">
	<Name>FinalLevel</Name>
	<ParameterType>integer</ParameterType>	
	<SkosConcept>http://example.edu/skos/finalLevell</SkosConcept>	
	<Unit>None</Unit>
	<Dimension xsi:type="AtomicConstantExpression" ConstantType="integer">
		<Constant>1</Constant>
	</Dimension>
</parameter>
\end{lstlisting}

\item $T$ is the thermodynamic temperature of the simulated medium and is expressed in Kelvin. The PDL {\it translation} for this sentence is:
\begin{lstlisting}[style=listXML]
<parameter dependency="required">
	<Name>Temperature</Name>
	<ParameterType>real</ParameterType>
	<SkosConcept>http://example.edu/skos/temperaturel</SkosConcept>
	<Unit>K</Unit>
	<Dimension xsi:type="AtomicConstantExpression" ConstantType="integer">
		<Constant>1</Constant>
	</Dimension>
</parameter>
\end{lstlisting}
\item $\rho$ is an electron density in $cm^{-3}$. The PDL version is:
\begin{lstlisting}[style=listXML]
<parameter dependency="required">
	<Name>Density</Name>
	<ParameterType>real</ParameterType>
	<SkosConcept>http://example.edu/skos/denisty</SkosConcept>
	<Unit>cm^-3</Unit>
	<Dimension xsi:type="AtomicConstantExpression" ConstantType="integer">
		<Constant>1</Constant>
	</Dimension>
</parameter>
\end{lstlisting}
\end{itemize} 
Even with this information, it is not guaranteed that users will be able to correctly use the service. Indeed, two constraints involve parameters. The first comes from the definition of $N_i$ and $N_f$: the condition
\begin{equation}\label{energyDifference}
(N_i-N_f)>1
\end{equation}
must always be satisfied. The second comes from the physical model implemented into the exposed code. The result has a physical meaning only if the Debey approximation hypothesis holds:
\begin{equation}\label{DebeyApprox}
\frac{9 \, \rho^{1/6}}{100 \, T^{1/2} }<1
\end{equation}
How to alert the user of these two constraints? A first solution consists in writing explanation (e.g. a code documentation) but it is not sure that users will read it. A more secure approach consists in writing checking algorithms. But this solution is time consuming, since you have to write {\it ad hoc} tests for every specific code. 
PDL answer this issues by providing a unified way for expressing the constraints. The PDL formulation of (\ref{energyDifference}) is 
\begin{lstlisting}[style=listXML]
<always>
	<Criterion xsi:type="Criterion">
 		<Expression xsi:type="AtomicParameterExpression">
			<parameterRef ParameterName="FinalLevel"/>
			<Operation operationType="MINUS">
				<Expression xsi:type="AtomicParameterExpression">
    					<parameterRef ParameterName="InitialLevel"/>
				</Expression>
                    	</Operation>
		</Expression>
                    <ConditionType xsi:type="ValueLargerThan" reached="true">
                    	<Value xsi:type="AtomicConstantExpression" ConstantType="real">
				<Constant>1</Constant>
                            </Value>
                    </ConditionType> 
 	</Criterion>
</always>
\end{lstlisting}
whereas the formulation for (\ref{DebeyApprox}) is
\begin{lstlisting}[style=listXML]
<always>
  <Criterion xsi:type="Criterion">
     <Expression xsi:type="AtomicConstantExpression" ConstantType="real">
           <Constant>0.09</Constant>
           <Operation operationType="MULTIPLY">
                <Expression xsi:type="AtomicParameterExpression">
                     <parameterRef ParameterName="Density"/>
                     <power xsi:type="AtomicConstantExpression" ConstantType="real">
                       <Constant>0.16666666</Constant>
                     </power>
                     <Operation operationType="DIVIDE">
                            <Expression xsi:type="AtomicParameterExpression">
                                <parameterRef ParameterName="Temperature"/>
                                 <power xsi:type="AtomicConstantExpression" ConstantType="real">
                                        <Constant>0.5</Constant>
                                 </power>
                           </Expression>
                     </Operation>
               </Expression>
         </Operation>
      </Expression>
      <ConditionType xsi:type="ValueSmallerThan" reached="false">
	<Value xsi:type="AtomicConstantExpression" ConstantType="real">
		<Constant>1</Constant>
	</Value>
      </ConditionType>
 </Criterion>
</always>
\end{lstlisting}
These two last pieces of XML (composing a wider  PDL description) are not intended for humans. They are parsed by the PDL framework for automatically generate the checking algorithms associated with the described constraints.\\

 The key point to retain is that PDL is simple for simple services is flexible and powerful enough for meeting description requirements coming with the most complex scientific codes (of course the associated description won't be simple).

\subsection{The PDL description of the example of equation (\ref{PDLExemplum01})}\label{Exemplum1XML}
The reader will find the xml file related to this example at the following URL:\\
\href{http://vo-param.googlecode.com/svn/trunk/model/documentation/PDL-Description_example01.xml}{http://vo-param.googlecode.com/svn/trunk/model/documentation/\\PDL-Description\_example01.xml}

\subsection{The PDL description of the example of equation (\ref{PDLExemplum02})}\label{Exemplum2XML}
The reader will find the xml file related to this example at the following URL:\\ 
\href{http://vo-param.googlecode.com/svn/trunk/model/documentation/PDL-Description_Example02.xml}{http://vo-param.googlecode.com/svn/trunk/model/documentation/\\PDL-Description\_Example02.xml}.

\subsection{The PDL XSD Schema}\label{pdlSchema}

\begin{lstlisting}[style=listXML]
<?xml version="1.0" encoding="UTF-8"?>
<xs:schema xmlns:xs="http://www.w3.org/2001/XMLSchema" xmlns:pm="http://www.ivoa.net/xml/PDL/v1.0"
   elementFormDefault="qualified" targetNamespace="http://www.ivoa.net/xml/PDL/v1.0">
   <!-- needs isActive property on group - need to be able to reference a group -->
   <xs:annotation>
      <xs:documentation> IVOA Description of the set of parameters for a service</xs:documentation>
   </xs:annotation>
   <xs:element name="Service">
      <xs:annotation>
         <xs:documentation> The base service description. A
            service in this context is simply some sort of process
            that has input parameters and produces output parameters.
         </xs:documentation>
      </xs:annotation>
      <xs:complexType>
         <xs:sequence>
            <xs:element name="ServiceId" type="xs:string" minOccurs="1" maxOccurs="1">
               <xs:annotation>
                  <xs:documentation>The ivoa identifier for the service</xs:documentation>
               </xs:annotation>
            </xs:element>
            <xs:element name="ServiceName" type="xs:string" minOccurs="1" maxOccurs="1"/>
            <xs:element name="Description" type="xs:string" minOccurs="1" maxOccurs="1"/>
            <xs:element name="Parameters" type="pm:Parameters" minOccurs="1" maxOccurs="1">
               <xs:annotation>
                  <xs:documentation>The list of all possible parameters both input and output parameters</xs:documentation>
               </xs:annotation>
            </xs:element>
            <xs:element name="Inputs" type="pm:ParameterGroup" minOccurs="1" maxOccurs="1">
               <xs:annotation>
                  <xs:documentation>The input parameters for a service.</xs:documentation>
               </xs:annotation>
            </xs:element>
            <xs:element name="Outputs" type="pm:ParameterGroup" minOccurs="1" maxOccurs="1">
               <xs:annotation>
                  <xs:documentation>The parameters output from a service.</xs:documentation>
               </xs:annotation>
            </xs:element>
         </xs:sequence>
      </xs:complexType>
      <!-- keys to ensure that parameter names are unique -->
      <xs:unique name="KeyName">
         <xs:selector xpath="./pm:ParameterList/pm:parameter"/>
         <xs:field xpath="pm:Name"/>
      </xs:unique>
      <xs:keyref name="expressionKeyref" refer="pm:KeyName">
         <xs:selector xpath=".//pm:parameterRef"/>
         <xs:field xpath="pm:parameterName"/>
      </xs:keyref>

   </xs:element>
   <xs:complexType name="Parameters">
      <xs:annotation>
         <xs:documentation>The list of possible parameters both input and output.</xs:documentation>
      </xs:annotation>
      <xs:sequence>
         <xs:element name="parameter" type="pm:SingleParameter" minOccurs="1" maxOccurs="unbounded">
         </xs:element>
      </xs:sequence>
   </xs:complexType>
   <xs:complexType name="ParameterReference">
      <xs:annotation>
         <xs:documentation>A reference to a parameter</xs:documentation>
      </xs:annotation>
      <xs:attribute name="ParameterName" type="xs:string">
         <xs:annotation>
            <xs:documentation>The name of the parameter being referred to.</xs:documentation>
         </xs:annotation>
      </xs:attribute>
   </xs:complexType>
   <xs:complexType name="Description">
      <xs:sequence>
         <xs:element name="humanReadableDescription" type="xs:string"/>
      </xs:sequence>
   </xs:complexType>

   <xs:simpleType name="ParameterDependency">
      <xs:annotation>
         <xs:documentation>The types that a parameter may have.</xs:documentation>
         <xs:documentation>
            Flag for saying if a parameter is required or optional   
         </xs:documentation>
      </xs:annotation>
      <xs:restriction base="xs:string">
         <xs:enumeration value="required">
            <xs:annotation>
               <xs:documentation>The parameter must be provided by user.</xs:documentation>
            </xs:annotation>
         </xs:enumeration>
         <xs:enumeration value="optional">
            <xs:annotation>
               <xs:documentation>The parameter is optional.</xs:documentation>
            </xs:annotation>
         </xs:enumeration>
      </xs:restriction>
   </xs:simpleType>

   <xs:simpleType name="ParameterType">
      <xs:annotation>
         <xs:documentation>The types that a parameter may have.</xs:documentation>
         <xs:documentation>
            Note that the types are made more specific by using the UCD attribute of the parameter definition. 
            In particular it is expected that a Parameter Model library would be able to recognise the more specific types associated with the following UCDs
            <ul>
              <li>pos - to provide a suitable widget for positions</li>
              <li>time - to provide suitable widgets for times and durations</li>
            </ul>
         </xs:documentation>
      </xs:annotation>
      <xs:restriction base="xs:string">
         <xs:enumeration value="boolean">
            <xs:annotation>
               <xs:documentation>A representation of a boolean - e.g. true/false</xs:documentation>
            </xs:annotation>
         </xs:enumeration>
         <xs:enumeration value="string">
            <xs:annotation>
               <xs:documentation>Data that can be interpreted as text.</xs:documentation>
            </xs:annotation>
         </xs:enumeration>
         <xs:enumeration value="integer"/>
         <xs:enumeration value="real"/>
         <xs:enumeration value="date"/>
      </xs:restriction>
   </xs:simpleType>

   <xs:simpleType name="FunctionType">
      <xs:restriction base="xs:string">
         <xs:enumeration value="size"/>
         <xs:enumeration value="abs"/>
         <xs:enumeration value="sin"/>
         <xs:enumeration value="cos"/>
         <xs:enumeration value="tan"/>
         <xs:enumeration value="asin"/>
         <xs:enumeration value="acos"/>
         <xs:enumeration value="atan"/>
         <xs:enumeration value="exp"/>
         <xs:enumeration value="log"/>
         <xs:enumeration value="sum"/>
         <xs:enumeration value="product"/>
      </xs:restriction>
   </xs:simpleType>

   <xs:simpleType name="OperationType">
      <xs:restriction base="xs:string">
         <xs:enumeration value="PLUS"/>
         <xs:enumeration value="MINUS"/>
         <xs:enumeration value="MULTIPLY"/>
         <xs:enumeration value="DIVIDE"/>
         <xs:enumeration value="SCALAR"/>
      </xs:restriction>
   </xs:simpleType>

   <xs:complexType name="SingleParameter">
      <xs:sequence>
         <xs:element name="Name" type="xs:string" minOccurs="1" maxOccurs="1"> </xs:element>
         <xs:element name="ParameterType" type="pm:ParameterType" minOccurs="1" maxOccurs="1"> </xs:element>
         <xs:element name="UCD" type="xs:string" maxOccurs="1" minOccurs="0"> </xs:element>
         <xs:element name="UType" type="xs:string" maxOccurs="1" minOccurs="0"/>
         <xs:element name="SkosConcept" type="xs:string" minOccurs="0" maxOccurs="1"/>
         <xs:element name="Unit" type="xs:string" minOccurs="0" maxOccurs="1"/>
         <xs:element name="Precision" type="pm:Expression" minOccurs="0" maxOccurs="1"/>
         <xs:element name="Dimension" type="pm:Expression" maxOccurs="1" minOccurs="1"/>
      </xs:sequence>
      <xs:attribute name="dependency" type="pm:ParameterDependency"> </xs:attribute>
   </xs:complexType>

   <xs:complexType name="ParameterGroup">
      <xs:annotation>
         <xs:documentation>A logical grouping of parameters</xs:documentation>
      </xs:annotation>
      <xs:sequence>
         <xs:element name="Name" type="xs:string" maxOccurs="1" minOccurs="1">
            <xs:annotation>
               <xs:documentation>The name of the parameter group which can be used for display</xs:documentation>
            </xs:annotation>
         </xs:element>
         <xs:element name="ParameterRef" type="pm:ParameterReference" minOccurs="0"
            maxOccurs="unbounded">
            <xs:annotation>
               <xs:documentation>The list of parameters that are in the group</xs:documentation>
            </xs:annotation>
         </xs:element>
         <xs:element name="ConstraintOnGroup" type="pm:ConstraintOnGroup" maxOccurs="1"
            minOccurs="0">
            <xs:annotation>
               <xs:documentation>The constraints on parameters in the group</xs:documentation>
            </xs:annotation>
         </xs:element>
         <xs:element name="ParameterGroup" type="pm:ParameterGroup" minOccurs="0"
            maxOccurs="unbounded">
            <xs:annotation>
               <xs:documentation>possibly nested parameter groups</xs:documentation>
            </xs:annotation>
         </xs:element>
         <xs:element name="Active" type="pm:WhenConditionalStatement" maxOccurs="1" minOccurs="0">
            <xs:annotation>
               <xs:documentation>It the goup active? i.e. should it be displayed - The default is yes if there is no active element, otherwise it is the result of the evaluation of the When conditional statement.</xs:documentation>
            </xs:annotation>
         </xs:element>
      </xs:sequence>
   </xs:complexType>

   <xs:complexType name="ConstraintOnGroup">
      <xs:annotation>
         <xs:documentation>The possible constraints on the parameters in a group</xs:documentation>
      </xs:annotation>
      <xs:sequence>
         <xs:element name="ConditionalStatement" type="pm:ConditionalStatement" minOccurs="0"
            maxOccurs="unbounded"/>
      </xs:sequence>
   </xs:complexType>

   <xs:complexType abstract="true" name="ConditionalStatement">
      <xs:sequence>
         <xs:element name="comment" type="xs:string" minOccurs="1" maxOccurs="1"/>
      </xs:sequence>
   </xs:complexType>

   <xs:complexType name="IfThenConditionalStatement">
      <xs:complexContent>
         <xs:extension base="pm:ConditionalStatement">
            <xs:sequence>
               <xs:element name="if" type="pm:If" minOccurs="1" maxOccurs="1"/>
               <xs:element name="then" type="pm:Then" minOccurs="1" maxOccurs="1"/>
            </xs:sequence>
         </xs:extension>
      </xs:complexContent>
   </xs:complexType>
   <xs:complexType name="AlwaysConditionalStatement">
      <xs:complexContent>
         <xs:extension base="pm:ConditionalStatement">
            <xs:sequence>
               <xs:element name="always" type="pm:Always" minOccurs="1" maxOccurs="1"/>
            </xs:sequence>
         </xs:extension>
      </xs:complexContent>
   </xs:complexType>

   <xs:complexType name="WhenConditionalStatement">
      <xs:annotation>
         <xs:documentation>
            A statement that has only a True or a False value
         </xs:documentation>
      </xs:annotation>
      <xs:complexContent>
         <xs:extension base="pm:ConditionalStatement">
            <xs:sequence>
               <xs:element name="when" type="pm:When"/>
            </xs:sequence>
         </xs:extension>
      </xs:complexContent>
   </xs:complexType>
   <xs:complexType abstract="true" name="LogicalConnector">
      <xs:sequence>
         <xs:element name="Criterion" type="pm:AbstractCriterion" minOccurs="1" maxOccurs="1"/>
      </xs:sequence>
   </xs:complexType>

   <xs:complexType name="And">
      <xs:complexContent>
         <xs:extension base="pm:LogicalConnector"/>
      </xs:complexContent>
   </xs:complexType>

   <xs:complexType name="Or">
      <xs:complexContent>
         <xs:extension base="pm:LogicalConnector"/>
      </xs:complexContent>
   </xs:complexType>

   <xs:complexType abstract="true" name="ConditionalClause">
      <xs:sequence>
         <xs:element name="Criterion" type="pm:AbstractCriterion" minOccurs="1" maxOccurs="1">
         </xs:element>
      </xs:sequence>
   </xs:complexType>

   <xs:complexType name="Always">
      <xs:complexContent>
         <xs:extension base="pm:ConditionalClause"/>
      </xs:complexContent>
   </xs:complexType>

   <xs:complexType name="If">
      <xs:complexContent>
         <xs:extension base="pm:ConditionalClause"/>
      </xs:complexContent>
   </xs:complexType>
   <xs:complexType name="Then">
      <xs:complexContent>
         <xs:extension base="pm:ConditionalClause"/>
      </xs:complexContent>
   </xs:complexType>
   <xs:complexType name="When">
      <xs:complexContent>
         <xs:extension base="pm:ConditionalClause"/>
      </xs:complexContent>
   </xs:complexType>
   <xs:complexType abstract="true" name="AbstractCondition"/>
   <xs:complexType name="IsNull">
      <xs:complexContent>
         <xs:extension base="pm:AbstractCondition"/>
      </xs:complexContent>
   </xs:complexType>
   <xs:complexType name="IsInteger">
      <xs:complexContent>
         <xs:extension base="pm:AbstractCondition"> </xs:extension>
      </xs:complexContent>
   </xs:complexType>
   <xs:complexType name="IsReal">
      <xs:complexContent>
         <xs:extension base="pm:AbstractCondition"> </xs:extension>
      </xs:complexContent>
   </xs:complexType>
   <xs:complexType name="BelongToSet">
      <xs:annotation>
         <xs:documentation>The value must belong to a set</xs:documentation>
      </xs:annotation>
      <xs:complexContent>
         <xs:extension base="pm:AbstractCondition">
            <xs:sequence>
               <xs:element name="Value" type="pm:Expression" minOccurs="1" maxOccurs="unbounded"/>
            </xs:sequence>
         </xs:extension>
      </xs:complexContent>
   </xs:complexType>
   <xs:complexType name="ValueLargerThan">
      <xs:complexContent>
         <xs:extension base="pm:AbstractCondition">
            <xs:sequence>
               <xs:element name="Value" type="pm:Expression" maxOccurs="1" minOccurs="1"/>
            </xs:sequence>
            <xs:attribute name="reached" type="xs:boolean"/>
         </xs:extension>
      </xs:complexContent>
   </xs:complexType>
   <xs:complexType name="ValueSmallerThan">
      <xs:complexContent>
         <xs:extension base="pm:AbstractCondition">
            <xs:sequence>
               <xs:element name="Value" type="pm:Expression" maxOccurs="1" minOccurs="1"/>
            </xs:sequence>
            <xs:attribute name="reached" type="xs:boolean"/>
         </xs:extension>
      </xs:complexContent>
   </xs:complexType>
   <xs:complexType name="ValueInRange">
      <xs:complexContent>
         <xs:extension base="pm:AbstractCondition">
            <xs:sequence>
               <xs:element name="Sup" type="pm:ValueSmallerThan" maxOccurs="1" minOccurs="1"/>
               <xs:element name="Inf" type="pm:ValueLargerThan" maxOccurs="1" minOccurs="1"/>
            </xs:sequence>
         </xs:extension>
      </xs:complexContent>
   </xs:complexType>
   <xs:complexType name="ValueDifferentFrom">
      <xs:complexContent>
         <xs:extension base="pm:AbstractCondition">
            <xs:sequence>
               <xs:element name="Value" type="pm:Expression" maxOccurs="1" minOccurs="1"/>
            </xs:sequence>
         </xs:extension>
      </xs:complexContent>
   </xs:complexType>
   <xs:complexType name="DefaultValue">
      <xs:complexContent>
         <xs:extension base="pm:AbstractCondition">
            <xs:sequence>
               <xs:element name="Value" type="pm:Expression" maxOccurs="1" minOccurs="1"/>
            </xs:sequence>
         </xs:extension>
      </xs:complexContent>
   </xs:complexType>

   <xs:complexType abstract="true" name="AbstractCriterion">
      <xs:sequence>
         <xs:element name="Expression" type="pm:Expression" minOccurs="1" maxOccurs="1"> </xs:element>
         <xs:element name="ConditionType" type="pm:AbstractCondition" minOccurs="1" maxOccurs="1"/>
         <xs:element name="LogicalConnector" type="pm:LogicalConnector" maxOccurs="1" minOccurs="0"
         />
      </xs:sequence>
   </xs:complexType>

   <xs:complexType name="Criterion">
      <xs:complexContent>
         <xs:extension base="pm:AbstractCriterion"> </xs:extension>
      </xs:complexContent>
   </xs:complexType>

   <xs:complexType name="ParenthesisCriterion">
      <xs:complexContent>
         <xs:extension base="pm:AbstractCriterion">
            <xs:sequence>
               <xs:element name="ExternalLogicalConnector" type="pm:LogicalConnector" maxOccurs="1"
                  minOccurs="0"/>
            </xs:sequence>
         </xs:extension>
      </xs:complexContent>
   </xs:complexType>

   <xs:complexType name="Function">
      <xs:complexContent>
         <xs:extension base="pm:Expression">
            <xs:sequence>
               <xs:element name="expression" type="pm:Expression"/>
            </xs:sequence>
            <xs:attribute name="functionName" type="pm:FunctionType"/>
         </xs:extension>
      </xs:complexContent>
   </xs:complexType>
   <xs:complexType name="Operation">
      <xs:sequence>
         <xs:element name="expression" type="pm:Expression" maxOccurs="1" minOccurs="1"/>
      </xs:sequence>
      <xs:attribute name="operationType" type="pm:OperationType"> </xs:attribute>
   </xs:complexType>
   <xs:complexType abstract="true" name="Expression"> </xs:complexType>
   <xs:complexType name="ParenthesisContent">
      <xs:complexContent>
         <xs:extension base="pm:Expression">
            <xs:sequence>
               <xs:element name="expression" type="pm:Expression" minOccurs="1" maxOccurs="1"/>
               <xs:element name="power" type="pm:Expression" maxOccurs="1" minOccurs="0"/>
               <xs:element name="Operation" type="pm:Operation" maxOccurs="1" minOccurs="0"/>
            </xs:sequence>
         </xs:extension>
      </xs:complexContent>
   </xs:complexType>
   <xs:complexType name="AtomicParameterExpression">
      <xs:complexContent>
         <xs:extension base="pm:Expression">
            <xs:sequence>
               <xs:element name="parameterRef" type="pm:ParameterReference" maxOccurs="1"
                  minOccurs="1"> </xs:element>
               <xs:element name="power" type="pm:Expression" maxOccurs="1" minOccurs="0"/>
               <xs:element name="Operation" type="pm:Operation" maxOccurs="1" minOccurs="0"/>
            </xs:sequence>
         </xs:extension>
      </xs:complexContent>
   </xs:complexType>
   <xs:complexType name="AtomicConstantExpression">
      <xs:complexContent>
         <xs:extension base="pm:Expression">
            <xs:sequence>
               <xs:element name="Constant" type="xs:string" maxOccurs="unbounded" minOccurs="1"/>
               <xs:element name="power" type="pm:Expression" maxOccurs="1" minOccurs="0"/>
               <xs:element name="Operation" type="pm:Operation" maxOccurs="1" minOccurs="0"/>
            </xs:sequence>
            <xs:attribute name="ConstantType" type="pm:ParameterType"/>
         </xs:extension>
      </xs:complexContent>
   </xs:complexType>
   <xs:complexType name="FunctionExpression">
      <xs:complexContent>
         <xs:extension base="pm:Expression">
            <xs:sequence>
               <xs:element name="Function" type="pm:Function" maxOccurs="1" minOccurs="1"/>
               <xs:element name="Power" type="pm:Expression" maxOccurs="1" minOccurs="0"/>
               <xs:element name="Operation" type="pm:Operation" maxOccurs="1" minOccurs="0"/>
            </xs:sequence>
         </xs:extension>
      </xs:complexContent>
   </xs:complexType>
</xs:schema>

\end{lstlisting}


\begin{thebibliography}{10} 

\bibitem{WSDL}{\sc R. Chinnici, J.J Moreau, A. Ryman, S. Weerawarana.}{\em Web Services Description Language (WSDL) Version 2.0 Part 1: Core Language}, W3C Recommendation 26 June 2007.

\bibitem{WADL}{\sc M. Hadley.}{\em Web Application Description Language}, W3C Member Submission 31 August 2009.

\bibitem{Babel1}{\sc Thomas G. W. Epperly, Gary Kumfert, Tamara Dahlgren, Dietmar Ebner, Jim Leek, Adrian Prantl, Scott Kohn.}{\em High-performance language interoperability for scientific computing through Babel}, 2011, International Journal of High-performance Computing Applications (IJHPCA), LLNL-JRNL-465223, DOI: 10.1177/1094342011414036

\bibitem{Babel2}{\sc Gary Kumfert, David E. Bernholdt, Thomas Epperly, James Kohl, Lois Curfman McInnes, Steven Parker, and Jaideep Ray.} {\em How the Common Component Architecture Advances Computational Science Proceedings of Scientific Discovery through Advanced Computing} (SciDAC 2006), June 2006, in J. Phys.: Conf. Series. (also LLNL Tech Report UCRL-CONF-222279)

\bibitem{Babel3}{\sc Benjamin A. Allan, Robert Armstrong, David E. Bernholdt, Felipe Bertrand, Kenneth Chiu, Tamara L. Dahlgren, Kostadin Damevski, Wael R. Elwasif, Thomas G. W. Epperly, Madhusudhan Govindaraju, Daniel S. Katz, James A. Kohl, Manoj Krishnan, Gary Kumfert, J. Walter Larson, Sophia Lefantzi, Michael J. Lewis, Allen D. Malony, Lois C. Mclnnes, Jarek Nieplocha, Boyana Norris, Steven G. Parker, Jaideep Ray, Sameer Shende, Theresa L. Windus, Shujia Zhou}. {\em A Component Architecture for High-Performance Scientific Computing}. Int. J. High Perform. Comput. Appl. 20(2). May 2006, pp. 163-202.
 
\bibitem{Taverna1}{\sc J. Sroka, J. Hidders, P. Missier, and C. Goble}. {\em A formal semantics for the Taverna 2 workflow model}, Journal of Computer and System Sciences, vol. 76, iss. 6, pp. 490-508, 2009.
 
\bibitem{Taverna2}{\sc D. Hull, K. Wolstencroft, R. Stevens, C. Goble, M. Pocock, P. Li, and T. Oinn}, {\em Taverna: a tool for building and running workflows of services}., Nucleic Acids Research, vol. 34, iss. Web Server issue, pp. 729-732, 2006. 

\bibitem{Osgi1}{\sc The Osgi Alliance.}{\em Osgi service platform, release 3}, Ios Press 2003, ISBN: 1586033115

\bibitem{Opalm1}{\sc S. Buis, A. Piacentini, D. D�clat}.{\em PALM: A Computational framework for assembling high performance computing applications}, Concurrency Computat.: Pract. Exper., Vol. 18(2), 2006, 247-262

\bibitem{Opalm2}{\sc T. Lagarde, A. Piacentini, O. Thual}. {\em A new representation of data assimilation methods: the PALM flow charting approach}, Q.J.R.M.S.,  Vol. 127 , 2001, pp. 189-207
 
\bibitem{Opalm3}{\sc A. Piacentini}.{\em The PALM Group, PALM: A Dynamic Parallel Coupler}. Lecture Notes In Computer Science, High Performance Computing for Computational Science, Vol. 2565, 2003, pp. 479-492

\bibitem{GumTree1}{\sc T. Lam, N. Xiong, P. Hathaway, N. Hauser. }{\em GumTree Decoded}, in proceedings of ICNS 2007 Conference.
 
\bibitem{GumTree2}{\sc T. Lam, N. Hauser, A. G�tz, P. Hathaway, F. Franceschini, H. Rayner, L. Zhang}{\em GumTree - An Integrated Scientific Experiment Environment}, in proceedings of ICNS 2005 Conference. 

\bibitem{Skos1}{\sc A. J. G. Gray, N. Gray, I. Ounis}. {\em Finding Data Resources in a Virtual Observatory Using SKOS Vocabularies,} BNCOD 2008:189-192.
 
\bibitem{Skos2} {\sc A. J. G. Gray, N. Gray, A. Paul Millar, and I. Ounis}{\em Semantically Enabled Vocabularies in Astronomy}, Technical Report, University of Glasgow, December 2007 (http://www.cs.man.ac.uk/~graya/Publications/vocabulariesInAstronomy.pdf).
 
\bibitem{Ontology} {\sc D.  Oberle, N. Guarino and S Staab}. 
{\em What is an ontology?},  In: Handbook on Ontologies. Springer, 2nd edition, 2009. 

\bibitem{UCD}{\sc S. Derriere, N. Gray, R. Mann, A. Preite Martinez, J. McDowell, T. Mc Glynn, F. Ochsenbein, P. Osuna, G. Rixon, R. Williams}{\em An IVOA Standard for Unified Content Descriptors}, International Virtual Observatory Alliance Recommendation, August 2005 (http://www.ivoa.net/documents/latest/UCD.html).

\bibitem{Utype}{\sc F. Bonnarel, I. Chilingarian, M. Louys, A. Micol, A. Richards, J. McDowell.}{\em Utype list for the Characterisation Data Model}, IVOA Note 25 June 2007 (http://www.ivoa.net/documents/latest/UtypeListCharacterisationDM.html).

\bibitem{Unit}{\sc S. Derriere, N. Gray, M. Louys, J. McDowell, F. Ochsenbein, P. Osuna, A Richards, B. Rino, J. Salgado.} {\em Units in the VO}, IVOA Proposed Recommendation 29 April 2013 (http://www.ivoa.net/documents/VOUnits/).

\bibitem{UWS}{\sc P. Harrison, G. Rixon}{\em Universal Worker Service Pattern}{IVOA Recommendation 10 October 2010 (http://www.ivoa.net/documents/UWS/)}

\bibitem{AstroTaverna}{\sc Garrido, J., Soiland-Reyes, S., Ruiz, J. E., S�anchez, S.}{\em AstroTaverna:
Tool for Scientific Workflows in Astronomy.} Astrophysics Source Code Library, record ascl:1307.007. 2013.

\bibitem{REST} {\sc R. T. Fielding}{\em Architectural Styles and the Design of Network-based software Architecture.} PhD Thesis in Information and computer Science, University of California, Irvine.

\bibitem{Soap}{\sc H. Haas, O. Hurley, A. Karmarkar, J. Mischkinsky, M. Jones, L. THompson, R. Martin }{\em W3C Reccomandation} http://www.w3.org/TR/2007/REC-soap12-testcollection-20070427/

\end{thebibliography}
\end{document}